\documentclass[a4paper,10pt]{revtex4}
\usepackage{mathrsfs}
\usepackage{graphicx}
\usepackage{latexsym}
\usepackage{amsmath}
\usepackage{amssymb}
\usepackage{textcomp}
\usepackage{amsbsy}
\usepackage{graphics}
\usepackage{epstopdf}
\usepackage{color}

\allowdisplaybreaks[4]

\begin{document}

\tolerance=5000

\title{Non-singular Bounce Cosmology from Lagrange Multiplier $F(R)$ Gravity}
\author{Shin'ichi~Nojiri,$^{1,2}$\,\thanks{nojiri@gravity.phys.nagoya-u.ac.jp}
S.~D.~Odintsov,$^{3,4,6}$\,\thanks{odintsov@ieec.uab.es}
V.~K.~Oikonomou,$^{5,6,7}$\,\thanks{v.k.oikonomou1979@gmail.com}
Tanmoy~Paul$^8$\thanks{pul.tnmy9@gmail.com}} \affiliation{ $^{1)}$
Department of Physics, Nagoya University,
Nagoya 464-8602, Japan \\
$^{2)}$ Kobayashi-Maskawa Institute for the Origin of Particles
and the Universe, Nagoya University, Nagoya 464-8602, Japan \\
$^{3)}$ ICREA, Passeig Luis Companys, 23, 08010 Barcelona, Spain\\
$^{4)}$ Institute of Space Sciences (IEEC-CSIC) C. Can Magrans
s/n,
08193 Barcelona, Spain\\
$^{5)}$ Department of Physics, Aristotle University of
Thessaloniki, Thessaloniki 54124,
Greece\\
$^{6)}$ International Laboratory for Theoretical Cosmology, Tomsk
State University of Control Systems
and Radioelectronics (TUSUR), 634050 Tomsk, Russia\\
$^{7)}$ Tomsk State Pedagogical University, 634061 Tomsk, Russia\\
$^{8)}$ Department of Theoretical Physics,\\
Indian Association for the Cultivation of Science,\\
2A $\&$ 2B Raja S.C. Mullick Road,\\
Kolkata - 700 032, India }

\tolerance=5000

\begin{abstract}
In this work we study non-singular{ bounce cosmology} in the context
of the Lagrange multiplier generalized $F(R)$ gravity theory of
gravity. We specify our study by using a specific variant form of
the well known matter{ bounce cosmology}, with scale factor $a(t) =
\left(a_0t^2 + 1 \right)^n$, and we demonstrate that for $n < 1/2$, the
primordial curvature perturbations are generated deeply in the
contraction era. Particularly, we show explicitly that the
perturbation modes exit the horizon at a large negative time
during the contraction era, which in turn makes the
``low-curvature'' regime, the era for which the calculations of
observational indices related to the primordial power spectrum can
be considered reliable. Using the reconstruction techniques for
the Lagrange multiplier $F(R)$ gravity, we construct the form of
effective $F(R)$ gravity that can realize such a cosmological
evolution, and we determine the power spectrum of the primordial
curvature perturbations. Accordingly, we calculate the spectral
index of the primordial curvature perturbations and the
tensor-to-scalar ratio, and we confront these with the latest
observational data. We also address the issue of stability of the
primordial metric perturbations, and to this end, we determine the
form of $F(R)$ which realizes the non-singular cosmology for the
whole range of cosmic time $-\infty < t < \infty$, by solving the
Friedmann equations without the ``low-curvature'' approximation.
This study is performed numerically though, due to the high
complexity of the resulting differential equations. By using this
numerical solution, we show that the stability is achieved for the
same range of values of the free parameters that guarantee the
phenomenological viability of the model. We also investigate the
energy conditions in the present context. The phenomenology of the
non-singular bounce is also studied in the context of a standard
$F(R)$ gravity. We find that the results obtained in the Lagrange
multiplier $F(R)$ gravity model have differences in comparison to
the standard $F(R)$ gravity model, where the observable indices
are not simultaneously compatible with the latest Planck results,
and also the standard $F(R)$ gravity model is plagued with
instabilities of the perturbation. These facts, clearly justify
the importance of the Lagrange multiplier field in making the
observational indices compatible with the Planck data and also in
removing the instabilities of the metric perturbations. Thereby,
the bounce with the aforementioned scale factor is adequately
described by the Lagrange multiplier $F(R)$ gravity, in comparison
to the standard $F(R)$ model.
\end{abstract}


\maketitle
\section{Introduction}

According to standard cosmology, the early Universe was dense and
hot and it seems that the timelike geodesics have a focal point or
a focal bundle, which in most cases is assumed that it leads to
the Big Bang primordial singularity. The Big Bang singularity is
an assumption or a direct consequence of classical equations of
motion, however this singularity cannot be perceived as a point in
spacetime, but as an initial spacelike three dimensional
hypersurface, due to the fact that if it was a point, this would
lead to infinitely causally disconnected regions in the Universe.
Apart from these theoretical conceptual problems, it is known that
any singularity in a classical Universe must be dressed inside a
horizon. It is possible that the Big Bang singularity which is a
past spacelike singularity, is just a manifestation of the
underlying quantum theory of gravity, as in classical
electrodynamics Coulomb potential singularities at the origin of
the potential, which are resolved in the context of quantum
electrodynamics. Having these in mind, it is apparent that the
theories which conceptually lead to a Big Bang singularity, are
haunted by the above conceptual problems.

An alternative theory that is free from the above problems, is the
so-called Big Bounce theory, or { bouncing cosmology} in general
\cite{Brandenberger:2012zb,Brandenberger:2016vhg,Battefeld:2014uga,Novello:2008ra,Cai:2014bea,deHaro:2015wda,Lehners:2011kr,Lehners:2008vx,
Cheung:2016wik,Cai:2016hea,Cattoen:2005dx,Mukhanov:1991zn,Li:2014era,Brizuela:2009nk,Cai:2013kja,Quintin:2014oea,Cai:2013vm,Poplawski:2011jz,Koehn:2015vvy,Odintsov:2015zza,Nojiri:2016ygo,Oikonomou:2015qha,Odintsov:2015ynk,Koehn:2013upa,Battarra:2014kga,Martin:2001ue,Khoury:2001wf,Buchbinder:2007ad,Brown:2004cs,Hackworth:2004xb,Nojiri:2006ww,Johnson:2011aa,Peter:2002cn,Gasperini:2003pb,Creminelli:2004jg,Lehners:2015mra,Mielczarek:2010ga,Lehners:2013cka,Cai:2014xxa,Cai:2007qw,Cai:2010zma,Avelino:2012ue,Barrow:2004ad,Haro:2015zda,Elizalde:2014uba,Das:2017jrl}.
{ Bouncing cosmologies} are free from primordial initial
singularities, since the Universe initially contracts until it
reaches a minimal size, and then bounces off at a specific cosmic
time instance and starts to expand again. This process can be
repeated for an infinite number of times, this is why sometimes
{ bouncing cosmologies} are also known as cyclic cosmologies.
{ Bounce cosmology} is also appealing since it is derived as a cosmological
solution from Loop Quantum Cosmology theory
\cite{Laguna:2006wr,Corichi:2007am,Bojowald:2008pu,Singh:2006im,
Date:2004fj,deHaro:2012xj,Cianfrani:2010ji,Cai:2014zga,
Mielczarek:2008zz,Mielczarek:2008zv,Diener:2014mia,Haro:2015oqa,
Zhang:2011qq,Zhang:2011vi,Cai:2014jla,WilsonEwing:2012pu}.

Among various bouncing models proposed over the last several
years, the matter bounce scenario
\cite{deHaro:2015wda,Finelli:2001sr,Quintin:2014oea,Cai:2011ci,
Haro:2015zta,Cai:2011zx,Cai:2013kja,
Haro:2014wha,Brandenberger:2009yt,deHaro:2014kxa,Odintsov:2014gea,
Qiu:2010ch,Oikonomou:2014jua,Bamba:2012ka,deHaro:2012xj,
WilsonEwing:2012pu} earned special attention, since it can provide
a nearly scale invariant power spectrum of primordial curvature
perturbations. The matter bounce scenario is essentially
characterized by the Universe evolved through a nearly matter
dominated epoch at very early times in the contracting phase, in
order to obtain an approximately scale invariant power spectrum,
and gradually evolves towards a bounce where all the parts of the
Universe become in causal contact \cite{Amoros:2013nxa}, solving
the horizon problem. After it bounces off, it enters a regular
expanding phase, in which it matches the behavior of the standard
Big Bang cosmology. However in order to obtain a viable matter
bounce scenario, it is expected that the underlying model is
consistent with various observational constraints that are put by
the latest Planck data. Moreover there are several conceptual
issues that are not clear in the framework of matter bounce
scenario. Firstly, in an exact matter bounce scenario,
materialized by using a single scalar field model, the power
spectrum is exactly scale invariant, which is in tension with the
observational constraints. The inconsistency of spectral index in
the context of matter bounce scenario was also confirmed in
\cite{Odintsov:2014gea} from a slightly different point of view.
Secondly, according to the Planck 2018 data, the running of the
spectral index is constrained to be $-0.0085 \pm 0.0073$. However,
for the single scalar field matter bounce scenario model, the
running of the index becomes zero and hence does not comply with
the observations. At this stage it deserves mentioning that the
running of the spectral index is still not a parameter of the
standard model of cosmology. In other words, $\alpha_s$ is
consistent with the value $0$. Indeed, one cannot say that
$\alpha_s$ is different than zero by much more than $1\sigma$. All
this to say that it is a little harsh to confirm that a model does
not comply with observations when it is within $2\sigma$. At most,
predicting $\alpha_s = 0$ could be in slight tension ($<2\sigma$)
with observations. Predicting running in excess of the measured
value (by more than $2\sigma$ for instance) is a bigger issue.
Thirdly, in the simplest model of matter bounce scenario, the
amplitude of tensor fluctuation is comparable to that of curvature
perturbation and thus the value of tensor-to-scalar ratio is of
the order $\sim \mathcal{O}(1)$, which is in conflict with the
Planck constraints. However, in a quasi-matter bounce scenario
(instead of an exact matter bounce), according to which the scale
factor of the Universe evolves as $t^{\frac{2}{3(1+w)}}$ (with $w
\neq 0$), deeply in the contracting era, it is possible to recover
the consistency of spectral index and the running index even in a
single scalar field model, but the tensor-to-scalar ratio is still
problematic. Moreover, in the context of standard $F(R)$ gravity,
neither matter bounce nor quasi-matter bounce are consistent with
the Planck data, as we will demonstrate at a later section.

Motivated by the above arguments, we shall consider a variant
non-singular bounce with scale factor $a(t) = (a_0t^2 + 1)^n$ in
the context of the Lagrange multiplier $F(R)$ theory of gravity
\cite{Nojiri:2017ygt} and try to explore whether the matter bounce
($n = 1/3$) or the quasi-matter bounce ($n\sim \mathcal{O}(1/3)$)
scenario is viable in such a generalized $F(R)$ gravity
\cite{Nojiri:2017ncd,Nojiri:2010wj} framework. Our discussions are
extended to investigate the stability conditions of the primordial
metric perturbations and the energy conditions in the present
context. We further study the phenomenology of the aforementioned
bouncing model in the context of standard $F(R)$ gravity. By
comparing the results obtained from the Lagrange multiplier $F(R)$
gravity with that of the standard $F(R)$ gravity model, we
establish the importance of the Lagrange multiplier field from
various perspectives.

The paper is organized as follows: in section \ref{sec_model}, we
briefly discuss the generalized $F(R)$ gravity model in the
presence of a Lagrange multiplier term. Sections
\ref{sec_reconstruction}, \ref{sec_perturbation}, and
\ref{sec_energy} are devoted to the explicit calculation of the
power spectrum, the observational indices, the stability
conditions of the primordial perturbations and the investigation
of the energy conditions in the Lagrange multiplier $F(R)$ gravity
model. Section \ref{sec_standard_F(R)} is devoted on the
realization of the bouncing model under study with standard $F(R)$
gravity, and its comparison with that of the Lagrange multiplier
$F(R)$ gravity model. The conclusions follow at the end of the
paper.

\section{Essential Features of Lagrange Multiplier $F(R)$ Gravity}\label{sec_model}

Let us briefly recall the formalism of the Lagrange multiplier
$F(R)$ gravity developed in Ref.~\cite{Nojiri:2017ygt}. The action
of the model is,
\begin{align}
S = \frac{1}{2\kappa^2} \int d^4x \sqrt{-g}\left[F(R)
+ \lambda\left(\partial_{\mu}\Phi\partial^{\mu}\Phi
+ G(R)\right)\right]\, ,
\label{action1}
\end{align}
where $\kappa^2 = \frac{1}{M^2}$ with $M$ be the four dimensional
Planck mass $\sim 10^{19}$ GeV. Here, $F(R)$ and $G(R)$ are two
differentiable functions of the Ricci scalar $R$, $\Phi$ is a
scalar field with a self coupling kinetic term and the coupling is
determined by the function $\lambda$, known as the Lagrange
multiplier, in the action (\ref{action1}). It was shown in
\cite{Nojiri:2017ygt} that such variant theory of $F(R)$ gravity
with the Lagrange multiplier term is free of ghosts. The variation
of the action with respect to the function $\lambda$ and with
respect to the scalar field $\Phi$ lead to the following equations
of motion,
\begin{align}
\partial_{\mu}\Phi\partial^{\mu}\Phi + G(R)=0\, , \quad
\nabla_{\mu}\left(\lambda\partial^{\mu}\Phi\right)=0\, .
\label{eqn scalar}
\end{align}
On the other hand, by varying the action with respect to the
metric tensor $g^{\mu\nu}$, we obtain,
\begin{align}
\frac{1}{2}F(R)g_{\mu\nu} - (F'(R) + \lambda G'(R))R_{\mu\nu}
 - \lambda \partial_{\mu}\Phi\partial_{\nu}\Phi
+ \left(\nabla_{\mu}\nabla_{\nu} - g_{\mu\nu}\nabla^2\right)
\left(F'(R) + \lambda G'(R)\right) = 0\, .
\label{eqn gravity}
\end{align}
As we are interested in cosmological scenario in the present
context, we shall assume that the background geometry is described
by a flat Friedman-Robertson-Walker (FRW) metric,
\begin{align}
ds^2 = -dt^2 + a(t)^2\left(dx^2 + dy^2 + dz^2\right)\, ,
\label{metric}
\end{align}
where $a(t)$ is the scale factor of the Universe. As it is evident
from Eq.~(\ref{metric}), the Universe is considered to be
homogeneous and isotropic and thus the function $\lambda$ and the
scalar field are taken as functions of the cosmic time $t$. In
effect of the metric given in Eq.~(\ref{metric}), the field
equations in (\ref{eqn scalar}) take the
following form,
\begin{align}
 -\dot{\Phi}^2 + G(R) = 0\, , \quad
\frac{d}{dt}\left(a^3\lambda\dot{\Phi}\right) = 0\, ,
 \label{FRW eqn1}
\end{align}
which can be solved as,
\begin{align}
\dot{\Phi} = \sqrt{G(R)} \, , \quad
a^3\lambda\dot{\Phi} = E\, ,
\label{FRW solution1}
\end{align}
with $E$ being a constant of the integration. Using these solutions
along with the FRW metric, we obtain the temporal and spatial
component of Eq.~(\ref{eqn gravity}) as follows,
\begin{align}
 -\frac{F(R)}{2} + 3(\dot{H} + H^2)\left(F'(R) + \frac{EG'}{a^3\sqrt{G}}\right)
 - \frac{E\sqrt{G}}{a^3}
 - 3H \frac{d}{dt}\left(F'(R) + \frac{EG'}{a^3\sqrt{G}}\right) = 0\, ,
 \label{FRW eqn 2}
\end{align}
and
\begin{align}
\frac{F(R)}{2} - (\dot{H} + 3H^2)\left(F'(R) + \frac{EG'}{a^3\sqrt{G}}\right)
+ \left(\frac{d^2}{dt^2} + 2H\frac{d}{dt}\right)\left(F'(R) + \frac{EG'}{a^3\sqrt{G}}\right)
= 0\, ,
\label{FRW eqn3}
\end{align}
respectively, where $H(t) = \frac{\dot{a}}{a}$ is the Hubble rate.
It may be noticed that for $E = 0$, the gravitational equations
become identical with those of standard $F(R)$ gravity. Having the
equations at hand, our next task is to reconstruct the forms of
$F(R)$ and $G(R)$ that can realize a { bouncing cosmology} of
specific form, which is the subject of the next section.

\section{Realization of bouncing cosmology}\label{sec_reconstruction}

In the present section, we shall investigate which functional
forms of $F(R)$ and $G(R)$ can realize a bouncing Universe
cosmological scenario, with the following scale factor,
\begin{align}
a(t) = \left(a_0t^2 + 1 \right)^n\, ,
\label{scale factor}
\end{align}
where $a_0$ and $n$ are the model free parameters,  with $a_0$
having mass dimension [+2], while $n$ is dimensionless. The
Universe's evolution in a general { bouncing cosmology}, consists of
two eras, an era of contraction and an era of expansion. It is
obvious that the above scale factor describes a contracting era
for the Universe, when $t\to -\infty$, then the Universe reaches a
bouncing point, at $t=0$ at which the Universe has a minimal size,
and then the Universe starts to expand again, for cosmic times
$t>0$. Hence, the Universe in this scenario never develops a
crushing type Big Bang singularity. It may be mentioned that for
$n=1/3$, the scale factor describes the matter bounce scenario.
Eq.~(\ref{scale factor}) leads to the following Hubble rate
and its first derivative,
\begin{align}
H(t) = \frac{2nt}{t^2 + 1/a_0 }\, , \quad
\dot{H}(t) = -2n\frac{t^2 - 1/a_0}{\left(t^2 +1/a_0\right)^2}\, .
\end{align}
With the help of the above expressions, the Ricci scalar is found
to be,
\begin{align}
R(t)=12H^2 + 6\dot{H}
=12n \left[\frac{(4n-1)t^2 + 1/a_0}{\left(t^2 + 1/a_0\right)^2}\right] \, .
\label{ricci scalar}
\end{align}
Using Eq.~(\ref{ricci scalar}), one can determine the cosmic time
as a function of the Ricci scalar, that is the function $t =
t(R)$. As a result, the Hubble rate and its first derivative can
be expressed in terms of $R$ (however this statement holds for all
analytic functions of $t$) and also the differential operator
$\frac{d}{dt}$ can be written as $\frac{d}{dt} = \dot{R}(R)
\frac{d}{dR}$. By plugging the resulting expressions in
Eqs.~(\ref{FRW eqn 2}) and (\ref{FRW eqn3}), we obtain
differential equations which determine the functional form of
$F(R)$, $G(R)$ fully in terms of $R$, and by solving those
differential equations, the forms of $F(R)$ and $G(R)$ can be
found. However the differential equations will become cumbersome
to provide analytic solutions and thus we consider the
low-curvature limit of the theory, that is, $\frac{R}{a_0} \ll 1$
for the purpose of reconstruction. This approximation will prove
to be quite useful since as we show in a later section, the
primordial perturbations of the matter bounce scenario are
generated deeply in the contraction era, at $t\to -\infty$ ($t^2
\gg 1/a_0$), in which case $H\ll \sqrt{a_0}$ and therefore the
curvature is quite small ($\frac{R}{a_0} \ll 1$).

During the low-curvature regime (or at large negative time),
$R(t)$ can be written as $R(t) \sim \frac{12n(4n-1)}{t^2}$ from
Eq.~(\ref{ricci scalar}). This helps to express the scale factor,
the Hubble rate, its first derivative and the differential
operators $d/dt$, $d^2/dt^2$, in terms of  the Ricci scalar $R$ as
follows,
\begin{align}
a(R) = \frac{\left[12na_0^n(4n-1)\right]^n}{R^n} \, , \quad
H(R) = 2n\sqrt{\frac{R}{12n(4n-1)}} \, , \quad
\dot{H}(R)= -2n\sqrt{\frac{R}{12n(4n-1)}}\, ,
\end{align}
and
\begin{align}
 \frac{d}{dt}=-24n(4n-1)\left[\frac{R}{12n(4n-1)}\right]^{3/2} \frac{d}{dR}
\, , \quad
\frac{d^2}{dt^2}= \frac{1}{3n(4n-1)} \left[R^3\frac{d^2}{dR^2}
+ \frac{3}{2}R^2\frac{d}{dR}\right]\, ,
\label{low curvature quantites}
\end{align}
respectively. By plugging back these expressions to
Eqs.~(\ref{FRW eqn 2}) and (\ref{FRW eqn3}), and by introducing
$J(R) = F(R) + \frac{2E\sqrt{G(R)}}{a^3(R)}$, we get the following differential
equations,
\begin{align}
\frac{2}{(4n-1)}R^2 \frac{d^2J}{dR^2} - \frac{(1-2n)}{(4n-1)}R\frac{dJ}{dR} - J(R)
= 0\, ,
\label{low curvature eqn1}
\end{align}
and
\begin{align}
F(R) = \frac{(6n-1)}{3n(4n-1)}R\frac{dJ}{dR} - \frac{(3-4n)}{3n(4n-1)}R^2\frac{d^2J}{dR^2}
 - \frac{2}{3n(4n-1)}R^3\frac{d^3J}{dR^3}\, .
\label{low curvature eqn2}
\end{align}
Eq.~(\ref{low curvature eqn1}) has the following solution,
\begin{align}
J(R) = A R^{\rho} + B R^{\delta} \, ,
\label{low curvature sol1}
\end{align}
where $\rho = \frac{1}{4}\left[3 - 2n - \sqrt{1 + 4n(5+n)}\right]$
and $\delta = \frac{1}{4}\left[3 - 2n + \sqrt{1 + 4n(5+n)}\right]$
and also $A$ and $B$ are integration constants having mass dimension
$[2-2\rho]$ and $[2-2\delta]$, respectively. This solution of
$J(R)$ along with Eq.~(\ref{low curvature eqn2}) lead to the
following functional form of $F(R)$,
\begin{align}
F(R)=&A \left[\frac{(6n-1)}{3n(4n-1)}\rho - \frac{(3-4n)}{3n(4n-1)}\rho(\rho-1)
 - \frac{2}{3n(4n-1)}\rho(\rho-1)(\rho-2)\right]R^{\rho}\nonumber\\
&+B \left[\frac{(6n-1)}{3n(4n-1)}\delta - \frac{(3-4n)}{3n(4n-1)}\delta(\delta-1)
 - \frac{2}{3n(4n-1)}\delta(\delta-1)(\delta-2)\right]R^{\delta}\nonumber\\
=&C R^{\rho} + D R^{\delta}\, ,
\label{low curvature sol2}
\end{align}
where $C$ and $D$ are the corresponding coefficients of $R^{\rho}$
and $R^{\delta}$, respectively. With these solutions, the effective
$f(R)$ can be written as,
\begin{align}
f(R)= F(R) + \lambda G(R)
= \frac{1}{2}\left[J(R) + F(R)\right]
= \frac{1}{2}(A+C) R^{\rho} + \frac{1}{2}(B+D) R^{\delta}
\label{low curvature sol3}
\end{align}
where we use the solution of $\lambda(t) = \frac{E}{a^3\sqrt{G}}$.
Thus Eqs.~(\ref{low curvature sol1}), (\ref{low curvature sol2}),
and (\ref{low curvature sol3}) are the main results of the present
section. In the next section we address concretely the
cosmological perturbations issue and we shall confront the theory
with the observational data.


\section{Cosmological perturbation : Observable quantities and the stability condition}\label{sec_perturbation}

In this section we shall study the first order metric
perturbations of the theory at hand, following
Refs.~\cite{Hwang:2005hb,Noh:2001ia,Hwang:2002fp}, where the scalar and tensor
perturbations are calculated for various variants of higher
curvature gravity models. Scalar, vector and tensor perturbations
are decoupled, as in general relativity, so that we can focus our
attention to tensor and scalar perturbations
separately.

\subsection{Scalar Perturbations}

The scalar perturbation of FRW background metric is defined as
follows,
\begin{align}
ds^2 = -(1 + 2\Psi)dt^2 + a(t)^2(1 - 2\Psi)\delta_{ij}dx^{i}dx^{j}\, ,
\label{sp1}
\end{align}
where $\Psi(t,\vec{x})$ denotes the scalar perturbation. In
principle, perturbations should always be expressed in
terms of gauge invariant quantities, in our case the comoving
curvature perturbation defined as $\Re = \Psi - aHv$, where,
$v(t,\vec{x})$ is the velocity perturbation. However, we shall
work in the comoving gauge condition, where the velocity
perturbation is taken as zero, thus with such gauge fixing $\Re =
\Psi$. Thereby, we can work with the perturbed variable
$\Psi(t,\vec{x})$. The perturbed action up to $\Psi^2$ order is
\cite{Hwang:2005hb},
\begin{align}
\delta S_{\psi} = \int dt d^3\vec{x} a(t) z(t)^2\left[\dot{\Psi}^2
 - \frac{1}{a^2}\left(\partial_i\Psi\right)^2\right]\, ,
\label{sp2}
\end{align}
where $z(t)$ has the following expression,
\begin{align}
z(t) = \frac{a(t)}{\left(H(t)
+ \frac{1}{2f'(R)}\frac{df'(R)}{dt}\right)} \sqrt{\frac{2E\sqrt{G}}{a^3}
+ \frac{3}{2f'(R)}\left(\frac{df'(R)}{dt}\right)^2}\, .
\label{sp3}
\end{align}
It is evident from Eq.~(\ref{sp2}), that $c_s^2 = 1$, which
guarantees the absence of ghost modes or equivalently one may
argue that the model is free from gradient instability. Also
stability of the scalar perturbations is ensured if $z(t)^2 > 0$.
This stability issue must be checked for all cosmic times,
including the bouncing point, in which case the low-curvature
approximation does not hold true anymore, and it will be
thoroughly studied at a later section. Specifically, we will
examine the stability of perturbation by reconstructing $F(R)$ and
$G(R)$ beyond the low-curvature limit, numerically, due to the
complexity of the resulting differential equations when the
low-curvature approximation does not hold true anymore.

However at present, we concentrate on determining various
observable quantities and specifically, the spectral index of the
primordial curvature perturbations, the tensor-to-scalar ratio and
the running of the spectral index, which are eventually determined
at the time of horizon exit. For the scale factor we consider in
the present paper, the horizon exit occurs during the
low-curvature regime deeply in the contracting era. Thereby, for
the purpose of finding the observable parameters, the condition
$R/a_0 \ll 1$ stands as a viable approximation.

In the low-curvature limit, we determine various terms present in
the expression of $z(t)$ (see Eq.~(\ref{sp3})) as,
\begin{align}
\frac{a(t)}{\left(H(t) + \frac{1}{2f'(R)}\frac{df'(R)}{dt}\right)}
= \frac{a_0^n\left(12n(4n-1)\right)^{n+1/2}}{R^{n+1/2}}
\left[2n - \frac{(\rho-1)\left[1 + \frac{\delta(\delta-1)(B+D)}{\rho(\rho-1)(A+C)}
R^{\delta-\rho}\right]}
{\left[1 + \frac{\delta(B+D)}{\rho(A+C)}R^{\delta-\rho}\right]}\right]^{-1}\, ,
\nonumber
\end{align}
and
\begin{align}
\frac{2E\sqrt{G}}{a^3} + \frac{3}{2f'(R)}\left(\frac{df'(R)}{dt}\right)^2
= R^{\rho} \left[(A-C) \left(1 + \frac{(B-D)}{(A-C)}R^{\delta-\rho}\right)
+ \frac{\rho(A+C)(\rho-1)^2
\left[1 + \frac{\delta(\delta-1)(B+D)}{\rho(\rho-1)(A+C)}
R^{\delta-\rho}\right]^2} {4n(4n-1)\left[1
+ \frac{\delta(B+D)}{\rho(A+C)}R^{\delta-\rho}\right]}\right]\, .
\nonumber
\end{align}
Consequently $z(t)$ takes the following form,
\begin{align}
z(t) = \sqrt{3}a_0^n \left[12n(4n-1)\right]^n \frac{\sqrt{P(R)}}{Q(R)}
\frac{1}{R^{n + 1/2 - \rho/2}}\, ,
\label{sp4}
\end{align}
where $P(R)$ and $Q(R)$ are defined as follows,
\begin{align}
P(R) = \left[4n(4n-1)(A-C) \left(1 + \frac{(B-D)}{(A-C)}R^{\delta-\rho}\right)
+ \frac{\rho(A+C)(\rho-1)^2
\left[1 + \frac{\delta(\delta-1)(B+D)}{\rho(\rho-1)(A+C)}R^{\delta-\rho}\right]^2}
{\left[1 + \frac{\delta(B+D)}{\rho(A+C)}R^{\delta-\rho}\right]}\right]\, ,
\label{P}
\end{align}
and
\begin{align}
Q(R) = \left[2n - \frac{(\rho-1)\left[1 + \frac{\delta(\delta-1)(B+D)}{\rho(\rho-1)(A+C)}
R^{\delta-\rho}\right]}
{\left[1 + \frac{\delta(B+D)}{\rho(A+C)}R^{\delta-\rho}\right]}\right]\, .
\label{Q}
\end{align}

Before moving further, at this stage, we check whether $Q(R)$ goes
to zero or equivalently $z(t) \rightarrow \infty$ at some point of
time. This issue is known to occur in Horndeski theories, see for
example \cite{Ijjas:2017pei,Quintin:2015rta,Battarra:2014tga}. It
is important to examine because as we will show that the
Mukhanov-Sasaki equation (which is essential to determine the
observable quantities) has a term containing $1/z(t)$ and moreover
the Mukhanov variable ($v=z\Psi$) diverges at the point when
$z(t)$ goes to infinity. As mentioned earlier that the
perturbations generate in the low curvature regime deeply in the
contracting era and thus the above expression of $Q$ can be
simplified as follows:

\begin{eqnarray}
 Q(R) = (2n - \rho + 1) - \frac{\delta(\delta - \rho)(B+D)}{\rho(A+C)}R^{\delta-\rho}
 \label{report1}
\end{eqnarray}
where $\rho = \frac{1}{4}\left[3 - 2n - \sqrt{1 +
4n(5+n)}\right]$, $\delta = \frac{1}{4}\left[3 - 2n + \sqrt{1 +
4n(5+n)}\right]$ and $A$, $B$ are two integration constants.
Further, recall, the explicit expressions of $C$ and $D$ (see
Eq.(\ref{low curvature sol2})) are given by
\begin{eqnarray}
 C = A\bigg[\frac{(6n-1)}{3n(4n-1)}\rho - \frac{(3-4n)}{3n(4n-1)}\rho(\rho-1) - \frac{2}{3n(4n-1)}\rho(\rho-1)(\rho-2)\bigg]\nonumber\\
 D = B\bigg[\frac{(6n-1)}{3n(4n-1)}\delta - \frac{(3-4n)}{3n(4n-1)}\delta(\delta-1) - \frac{2}{3n(4n-1)}\delta(\delta-1)(\delta-2)\bigg]
 \nonumber
\end{eqnarray}
Putting these expressions of $C$ and $D$ into Eq.(\ref{report1}),
\begin{eqnarray}
 Q(R) = (2n - \rho + 1) -
 \frac{B\delta(\delta - \rho)
 \bigg(1 + \frac{(6n-1)}{3n(4n-1)}\delta - \frac{(3-4n)}{3n(4n-1)}\delta(\delta-1) - \frac{2}{3n(4n-1)}\delta(\delta-1)(\delta-2)\bigg)}
 {A\rho\bigg(1 + \frac{(6n-1)}{3n(4n-1)}\rho - \frac{(3-4n)}{3n(4n-1)}\rho(\rho-1) - \frac{2}{3n(4n-1)}\rho(\rho-1)(\rho-2)\bigg)}R^{\delta-\rho}
 \label{report2}
\end{eqnarray}
Using the forms of $\rho$ and $\delta$ (in terms of n), it can be
checked that the above expression of $Q$ is a positive definite
quantity (or does not hit to the value zero) for $n >
\frac{1}{4}$. Moreover we will show in the later sections that the
observable quantities are compatible with Planck observations for
the parametric regime $0.27 \lesssim n \lesssim 0.40$ (i.e for $n
> 1/4$). Therefore $Q(t)$ does not hit to zero or equivalently
$z(t)$ does not tend to infinity for the parametric values which
are consistent with the Planck observations. It may be mentioned
that such non-divergent character of $z(t)$ has been investigated
here in the low curvature regime (or at large negative time where
the pertutbation modes are generated). Thus there remains the
possibility that $z(t)$ goes to infinity in the large curvature
regime near the bounce phase, however, this may not be a physical
issue and may be resolved by studying the perturbations near this
`singularity’ in the perturbation equations in another gauge.\\
Eq.~(\ref{sp2}) clearly indicates that $\Psi(t,\vec{x})$ is not
canonically normalized and to this end we introduce the well-known
Mukhanov-Sasaki variable as $v = z\Re$ ($= z\Psi$ as we are
working in the comoving gauge). The corresponding fourier mode of
the Mukhanov-Sasaki variable satisfies,
\begin{align}
\frac{d^2v_k}{d\tau^2} + \left(k^2 - \frac{1}{z(\tau)}\frac{d^2z}{d\tau^2}\right)v_k(\tau)
= 0 \, ,
\label{sp5}
\end{align}
where $\tau = \int dt/a(t)$ is the conformal time and $v_k(\tau)$
is the Fourier transformed variable of $v(t,\vec{x})$ for the
$k$th mode. Eq.~(\ref{sp5}) is quite hard to solve analytically in
general, since the function $z$ depends on the background
dynamics. However the equation can be solved analytically in the
regime $R/a_0 \ll 1$ as we now show. The conformal time ($\tau$)
is related to the cosmic time ($t$) as $\tau = \int
\frac{dt}{a(t)} = \frac{1}{a_0^n(1-2n)} t^{1-2n}$ for $n \neq
1/2$, however we will show that the observable quantities are
compatible with Planck data \cite{Akrami:2018odb} for $n < 1/2$
and thus we can safely work with the aforementioned expression of
$\tau = \tau(t)$. Using this, we can express the Ricci scalar as a
function of the conformal time,
\begin{align}
R(\tau)= \frac{12n(4n-1)}{t^2}
=\frac{12n(4n-1)}{\left[a_0^n(1-2n)\right]^{2/(1-2n)}}
\frac{1}{\tau^{2/(1-2n)}}\, .
\label{sp6}
\end{align}
Having this in mind, along with Eq.~(\ref{sp4}), we can express
$z$ in terms of $\tau$ as follows,
\begin{align}
z(\tau) = \sqrt{3}a_0^n \left[12n(4n-1)\right]^n \frac{\sqrt{P(\tau)}}{Q(\tau)}
\tau^{\frac{2n+1-\rho}{1-2n}}\, .
\label{sp67}
\end{align}
The above expression of $z = z(\tau)$ yields the expression of
$\frac{1}{z}\frac{d^2z}{d\tau^2}$, which is essential for the
Mukhanov equation,
\begin{align}
\frac{1}{z}\frac{d^2z}{d\tau^2}=&\frac{\xi(\xi-1)}{\tau^2}
\left[1 + \frac{2(\delta-\rho)}{(\xi-1)}R^{\delta-\rho} \right. \nonumber\\
&\left. \times \left(\frac{\delta(\rho-\delta)(B+D)}{\rho(A+C)(2n-\rho+1)}
+ \frac{\delta(1-\rho)(1+\rho-2\delta)(B+D) + 4(B-D)n + 16(B-D)n^2}
{\rho(1-\rho)^2(A+C) + 4(A-C)n + 16(A-C)n^2}\right)\right] \, ,
\label{sp7}
\end{align}
with $\xi = \frac{(2n+1-\rho)}{(1-2n)}$. Recall $\rho =
\frac{1}{4}\left[3 - 2n - \sqrt{1 + 4n(5+n)}\right]$ and $\delta =
\frac{1}{4}\left[3 - 2n + \sqrt{1 + 4n(5+n)}\right]$, which clearly
indicate that $\delta - \rho$ is a positive quantity. Thus the
term within parenthesis in Eq.~(\ref{sp7}) can be safely
considered to be small in the low-curvature regime $R/a_0 \ll 1$.
As a result, $\frac{1}{z}\frac{d^2z}{d\tau^2}$ becomes
proportional to $1/\tau^2$ i.e., $\frac{1}{z}\frac{d^2z}{d\tau^2} =
\sigma/\tau^2$ with,
\begin{align}
\sigma = \xi(\xi-1)&\left[1 + \frac{2(\delta-\rho)}{(r-1)}R^{\delta-\rho}
\right. \nonumber\\
& \left. \times \left(\frac{\delta(\rho-\delta)(B+D)}{\rho(A+C)(2n-\rho+1)}
+ \frac{\delta(1-\rho)(1+\rho-2\delta)(B+D) + 4(B-D)n + 16(B-D)n^2}
{\rho(1-\rho)^2(A+C) + 4(A-C)n + 16(A-C)n^2}\right)\right]\, ,
\label{spnew}
\end{align}
which is approximately a constant in the era, when the primordial
perturbation modes generate deep inside the Hubble radius. In
effect, and in conjunction with the fact that $c_s^2 = 1$, the
Mukhanov equation can be solved as follows,
\begin{align}
v(k,\tau) = \frac{\sqrt{\pi|\tau|}}{2} \left[c_1(k)H_{\nu}^{(1)}(k|\tau|) +
c_2(k)H_{\nu}^{(2)}(k|\tau|)\right]\, ,
\label{sp8}
\end{align}
with $\nu = \sqrt{\sigma + \frac{1}{4}}$ and $c_1$ and $c_2$ are
integration constants. Assuming the Bunch-Davies vacuum initially,
these integration constants become $c_1 = 0$ and $c_2 =1$,
respectively. Having the solution of $v_k(\tau)$ at hand, next we
proceed to evaluate the power spectrum (defined for the
Bunch-Davies vacuum state) corresponding to the $k$-th scalar
perturbation mode, which is defined as follows,
\begin{align}
P_{\Psi}(k,\tau) = \frac{k^3}{2\pi^2}\left|\Psi_k(\tau)\right|^2
= \frac{k^3}{2\pi^2}\left|\frac{v_k(\tau)}{z(\tau)}\right|^2\, .
\label{sp9}
\end{align}
In the superhorizon limit, using the mode solution in
Eq.~(\ref{sp8}), we have,
\begin{align}
P_{\Psi}(k,\tau) = \left[\frac{1}{2\pi}\frac{1}{z|\tau|}
\frac{\Gamma(\nu)}{\Gamma(3/2)}\right]^2 \left(\frac{k|\tau|}{2}\right)^{3 - 2\nu}\, .
\label{sp10}
\end{align}
By using Eq.~(\ref{sp10}), we can determine the observable
quantities like spectral index of the primordial curvature
perturbations and the running of spectral index. Before proceeding
to calculate these observable quantities, we will consider first
the tensor power spectrum, which is necessary for evaluating the
tensor-to-scalar ratio.


\subsection{Tensor Perturbations}

Let us now focus on the tensor perturbations, and the tensor
perturbation on the FRW metric background is defined as follows,
\begin{align}
ds^2 = -dt^2 + a(t)^2\left(\delta_{ij} + h_{ij}\right)dx^idx^j\, ,
\label{tp1}
\end{align}
where $h_{ij}(t,\vec{x})$ is the tensor perturbation. The tensor
perturbation is itself a gauge invariant quantity, and the tensor
perturbed action up to quadratic order is given by,
\begin{align}
\delta S_{h} = \int dt d^3\vec{x} a(t) z_T(t)^2\left[\dot{h_{ij}}\dot{h^{ij}}
 - \frac{1}{a^2}\left(\partial_kh_{ij}\right)^2\right]\, ,
\label{tp2}
\end{align}
where $z_T(t)$ is given by
\begin{align}
z_T(t) = a\sqrt{f'(R)}\, ,
\label{tp3}
\end{align}
Therefore, the speed of the tensor perturbation propagation is
$c_T^2 = 1$. Moreover the tensor perturbation is stable if the
condition $z_T^2 > 0$ is satisfied, and in a later section we
shall examine in detail whether this condition is satisfied.

Similar to scalar perturbation, the Mukhanov-Sasaki variable for
tensor perturbation is defined as $(v_T)_{ij} = z_T~h_{ij}$ which,
upon performing the Fourier transformation, satisfies the
following equation,
\begin{align}
\frac{d^2v_T(k,\tau)}{d\tau^2}
+ \left(k^2 - \frac{1}{z_T(\tau)}\frac{d^2z_T}{d\tau^2}\right)v_T(k,\tau)
= 0\, .
\label{tp4}
\end{align}
By using Eq.~(\ref{tp3}), along with the condition $R/a_0 \ll 1$,
we evaluate $z_T(\tau)$ and
$\frac{1}{z_T(\tau)}\frac{d^2z_T}{d\tau^2}$ and these read,
\begin{align}
z_T(\tau) = a_0^n \left[12n(4n-1)\right]^n S(\tau)
\tau^{\frac{2n+1-\rho}{1-2n}}\, ,
\label{tpnew}
\end{align}
and
\begin{align}
\frac{1}{z_T}\frac{d^2z_T}{d\tau^2} = \frac{\xi(\xi-1)}{\tau^2}
\left[1 - \frac{2\delta(\delta-\rho)(B+D)}{(r-1)\rho(A+C)}
R^{\delta-\rho}\right]\, ,
\label{tp6}
\end{align}
respectively, where { $S(R(\tau)) = \sqrt{\frac{\rho(A+C)}{2}}
\left[1 + \frac{\delta(B+D)}{\rho(A+C)}R^{\delta-\rho}\right]^{1/2}$}
and also we use $R=R(\tau)$ from Eq.~(\ref{sp6}). Due to the fact that
$\delta-\rho$ is positive, the variation of the term in the
parenthesis in Eq.~(\ref{tp6}), can be regarded to be small in the
low-curvature regime and thus
$\frac{1}{z_T}\frac{d^2z_T}{d\tau^2}$ becomes proportional to
$1/\tau^2$ that is $\frac{1}{z_T}\frac{d^2z_T}{d\tau^2} =
\sigma_T/\tau^2$, with
\begin{align}
\sigma_T = \xi(\xi-1) \left[1 - \frac{2\delta(\delta-\rho)(B+D)}{(r-1)\rho(A+C)}
R^{\delta-\rho}\right]\, ,
\label{tp9}
\end{align}
and recall $\xi = \frac{(2n+1-\rho)}{(1-2n)}$. The above
expressions yield the tensor power spectrum, defined with initial
state the Bunch-Davies vacuum, so we have,
\begin{align}
P_{h}(k,\tau) = 2\left[\frac{1}{2\pi}\frac{1}{z_T|\tau|}
\frac{\Gamma(\nu_T)}{\Gamma(3/2)}\right]^2 \left(\frac{k|\tau|}{2}
\right)^{3 - 2\nu_T}\, .
\label{tp10}
\end{align}
The factor $2$ arises due to the two polarization modes of the
gravity wave, and $\nu_T = \sqrt{\sigma_T + \frac{1}{4}}$ where
$\sigma_T$ is defined in Eq.~(\ref{tp9}).

\begin{figure}[!h]
\begin{center}
 \centering
 \includegraphics[width=3.5in,height=2.0in]{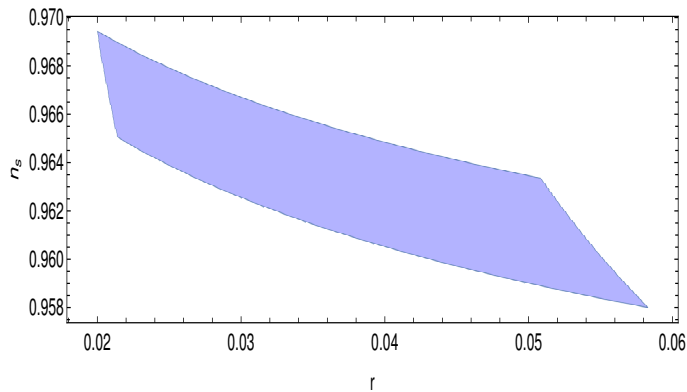}
 \caption{Parametric plot of $n_s$ vs $r$ for
 $0.01 \leq \frac{R_h}{a_0} \leq 0.07$ and $0.27 \lesssim n \lesssim 0.40$.}
 \label{plot1}
\end{center}
\end{figure}

Now we can explicitly confront the model at hand with the latest
Planck observational data \cite{Akrami:2018odb}, so we shall
calculate the spectral index of the primordial curvature
perturbations $n_s$ and the tensor-to-scalar ratio $r$, which are
defined as follows,
\begin{align}
n_s = 1 + \left. \frac{\partial \ln{P_{\Psi}}}{\partial
\ln{k}}\right|_{\tau=\tau_h} \, , \quad
r = \left. \frac{P_{h}(k,\tau)}{P_{\Psi}(k,\tau)}\right|_{\tau=\tau_h}\, .
\label{obs1}
\end{align}
Eqs.~(\ref{sp10}) and (\ref{tp10}) immediately lead to the
explicitly form of $n_s$ and $r$ as follows,
\begin{align}
n_s = 4 - \sqrt{1 + 4\sigma} \, , \quad
r = 2\left[\frac{z(\tau)}{z_T(\tau)}\right]^2_{\tau = \tau_h}\, ,
\label{obs2}
\end{align}
where $\sigma$, $z(\tau)$ and $z_T(\tau)$ are given in
Eqs.~(\ref{spnew}), (\ref{sp6}), and (\ref{tpnew}), respectively. As it
is evident from the above equations, $n_s$ and $r$ are evaluated
at the time of horizon exit, when $k=aH$, or equivalently at $\tau
= \tau_h$. It may be noticed that $n_s$ and $r$ depend on the
dimensionless parameters $\frac{R_h}{a_0}$ and $n$ with $R_h =
R(\tau_h)$. We can now directly confront the spectral index and
the tensor-to-scalar ratio with the Planck 2018 constraints
\cite{Akrami:2018odb}, which constrain the observational indices
as follows,
\begin{equation}
\label{planckconstraints}
n_s = 0.9649 \pm 0.0042\, , \quad r < 0.064\, .
\end{equation}
For the model at hand, $n_s$ and $r$ are within the Planck
constraints for the following ranges of parameter values: $0.01
\leq \frac{R_h}{a_0} \leq 0.07$ and $0.27 \lesssim n \lesssim
0.40$ and this behavior is depicted in Fig.~\ref{plot1}. The
viable range of $R_h/a_0$ is in agreement with the low-curvature
condition $R/a_0 \ll 1$ that we have considered in our
calculations. Moreover the range of the parameter $n$ clearly
indicates that the matter bounce scenario, for which $n=1/3$, is
well described by the generalized $F(R)$ gravity model with the
Lagrange multiplier term. At this stage it is worth mentioning
that in scalar-tensor theory (with an exponential scalar
potential), the matter bounce scenario is not consistent with
the Planck observations. Moreover the matter bounce scenario also does
not fit well even in the standard $F(R)$ gravity, as we also confirm
in a later section. However here, we show that in the presence of the
Lagrange multiplier generalized $F(R)$ gravity, the matter bounce
may be considered as a good bouncing model, which allows the
simultaneous compatibility of $n_s$ and $r$ with observations.\\
The results seem to indicate that $r$ can be suppressed below the
current observational bound. One can attribute this to the fact
that either scalar fluctuations have been enhanced or that tensor
fluctuations have been suppressed in comparison to the `standard'
model with Einstein gravity and matter contraction. So, we want to
explore which effect comes into play. For Einstein gravity with a
scalar field, the gravitational equations of motion turn out to
be,
\begin{eqnarray}
 H^2 = \frac{1}{3}\big[\frac{1}{2}\dot{\Phi}^2 + V(\Phi)\big]\nonumber\\
 2\dot{H} + 3H^2 + \frac{1}{2}\dot{\Phi}^2 - V(\Phi) = 0
 \label{ts_comparison1}
\end{eqnarray}

\begin{figure}[!h]
\begin{center}
 \centering
 \includegraphics[width=3.0in,height=2.0in]{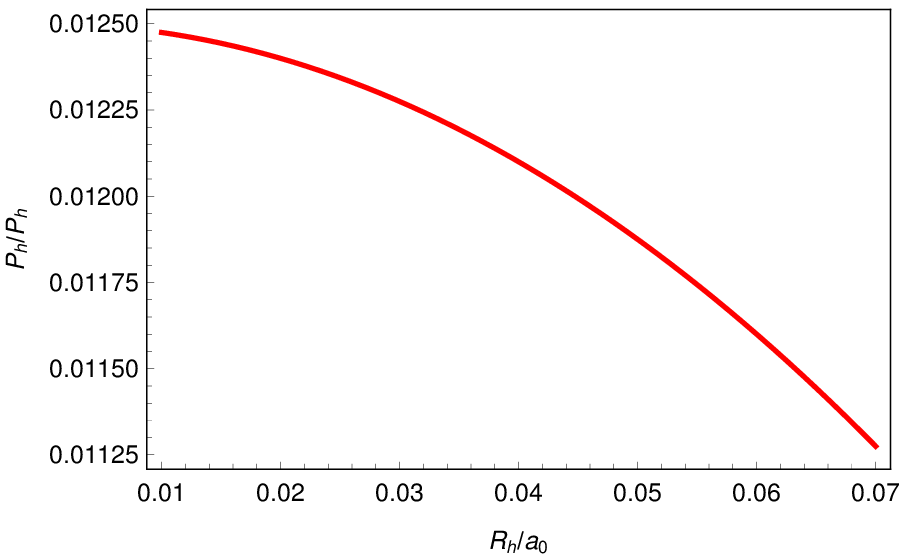}
 \includegraphics[width=3.0in,height=2.0in]{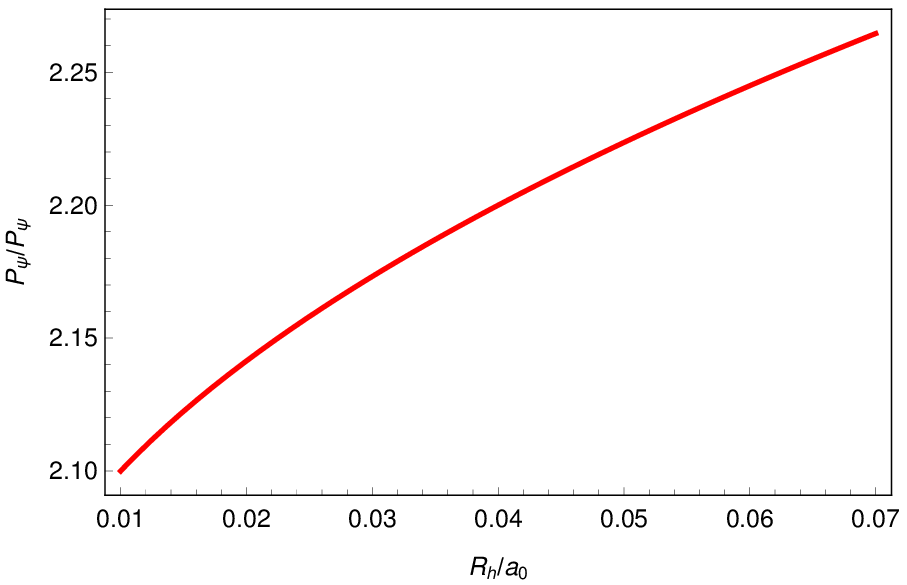}
 \caption{$Left~part$ : $P_h/\bar{P}_h$ vs. $R_h/a_0$ for the purpose of weak energy condition. $Right~part$ : $P_{\Psi}/\bar{P}_{\Psi}$ vs. $R_h/a_0$
 with $n = 1/3$.}
 \label{plottscomparison}
\end{center}
\end{figure}

\begin{figure}[!h]
\begin{center}
 \centering
 \includegraphics[width=3.5in,height=2.0in]{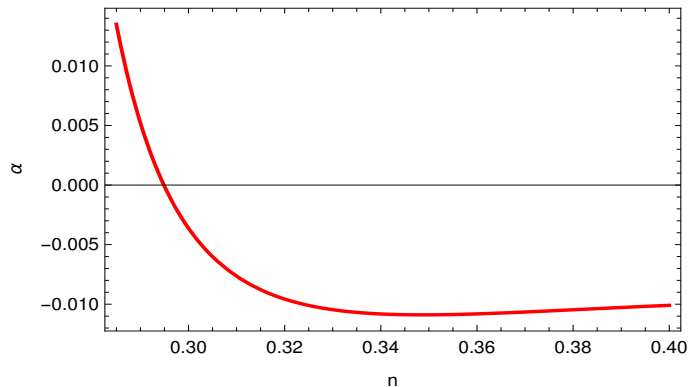}
 \caption{Parametric plot of $\alpha$ vs $n$ for
 $\frac{R_h}{a_0} = 0.05$ and $0.26 \lesssim n \lesssim 0.40$.}
 \label{plot2}
\end{center}
\end{figure}

Considering the scale factor $a(t) = (a_0t^2 + 1)^n$ along with the help of the above equations of motion, we get the scalar field dynamics and the Ricci
scalar as follows:
\begin{eqnarray}
 \dot{\Phi}(t)&=&\frac{2\sqrt{n}}{t}\nonumber\\
 R(t)&=&\frac{12n(4n-1)}{t^2}
 \label{ts_comparison2}
\end{eqnarray}
The above solutions are valid under the approximation $t^2 \gg 1/a_0$ which is a valid one as the perturbations generate
at large negative time deeply in the contracting era. Using the form of the scale factor, we obtain the conformal time in terms of the cosmic time
as $\tau \propto t^{1-2n}$. These expressions lead to the power spectrum for scalar and tensor perturbations in the case of the standard model
of Einstein gravity with matter contraction as follows:
\begin{eqnarray}
 \bar{P}_{\Psi}(k,\tau) = \left[\frac{1}{2\pi}\frac{1}{\bar{z}|\tau|}
\frac{\Gamma(\bar{\nu})}{\Gamma(3/2)}\right]^2 \left(\frac{k|\tau|}{2}\right)^{3 - 2\bar{\nu}}\, .
\label{ts_comparison3}
\end{eqnarray}
and
\begin{eqnarray}
 \bar{P}_{h}(k,\tau) = \frac{2}{M_{Pl}^2}\left[\frac{1}{2\pi}\frac{1}{\bar{z}_T|\tau|}
\frac{\Gamma(\bar{\nu}_T)}{\Gamma(3/2)}\right]^2 \left(\frac{k|\tau|}{2}
\right)^{3 - 2\bar{\nu}_T}\, .
\label{ts_comparison4}
\end{eqnarray}
respectively, where the quantities with bar denote the respective quantities in Einstein gravity and $M_{Pl}$ is the Planck mass.
Moreover the explicit expressions of the barred quantities
are following :
\begin{eqnarray}
 \bar{z}&=&\frac{a(t)\dot{\Phi}}{H}\bigg|_{t_h} = \frac{\big[12n(4n-1)\big]^n}{\sqrt{n}}\bigg(\frac{a_0}{R_h}\bigg)^n\nonumber\\
 \bar{z}_T&=&\frac{a(t)}{2}\bigg|_{t_h} = \frac{1}{2}\big[12n(4n-1)\big]^n \bigg(\frac{a_0}{R_h}\bigg)^n\nonumber\\
 \bar{\nu}&=&\bar{\nu}_T = \sqrt{\frac{2n(4n-1)}{(1-2n)^2} + \frac{1}{4}}
 \label{ts_comparison5}
\end{eqnarray}
where $t_h$ is the horizon crossing time, $R(t_h) = R_h$ and the
factor of $1/2$ in the expression of $\bar{z}_T$ ensures that the
perturbed action for tensor modes is truly canonically normalized
when written in terms of the Mukhanov variable. Therefore it is
clear that for matter or quasi-matter bounce in Einstein gravity,
the scalar and tensor power spectrums are comparable to each other
and thus the tensor to scalar ratio ($r =
\frac{\bar{P}_h}{\bar{P}_{\Psi}}$) becomes of the order unity.
However in the Lagrange multiplier F(R) gravity model, the tensor
to scalar ratio gets suppressed and matches with the Planck
constraints even for matter or quasi-matter bounce (as we showed
earlier). In order to compare the perturbations of Lagrange
multiplier F(R) gravity model with that of the standard model of
Einstein gravity, we give plots of the ratio of the respective
power spectrums i.e $P_h/\bar{P}_h$ and $P_{\Psi}/\bar{P}_{\Psi}$
in terms of the parameter $R_h/a_0$ with $n = 1/3$ (see Fig[\ref{plottscomparison}]).\\
To obtain the plots we use the horizon crossing relation $k = aH$.
Fig.[\ref{plottscomparison}] clearly demonstrates that the tensor
power spectrum gets suppressed in the Lagrange multiplier F(R)
model in comparison to Einstein gravity, while the scalar power
spectrums remain of same order in both the aforementioned models.
This leads to a suppressed tensor to scalar ratio in the Lagrange
multiplier F(R) gravity with respect to the standard Einstein
model. However the presence of Lagrange multiplier term may also
effect on the production of non-Gaussianities, as is also known
for $k$-essence theories \cite{Li:2016xjb} and Horndeski theories
\cite{Akama:2019qeh}. We hope to address this issue in a future
work.\\ Furthermore, the running of the spectral index is defined
as follows,
\begin{align}
 \alpha = \left. \frac{dn_s}{d\ln{k}} \right|_{\tau=\tau_h}\, ,
 \label{obs3}
\end{align}
and this is constrained by Planck 2018 results as $\alpha =
-0.0085 \pm 0.0073$. Thus, it is also important to calculate the
running of spectral index before concluding the viability of a
model. By using the expression of $\sigma$ (see Eq.~(\ref{spnew}))
and $R = R(\tau)$ (see Eq.~(\ref{sp6})), we get
\begin{align}
\alpha=&\frac{4\xi(\delta-\rho)^2}{\sqrt{1 + 4\xi(\xi-1)}}R_h^{\delta-\rho}
\nonumber\\
& \times \left(\frac{\delta(\rho-\delta)(B+D)}{\rho(A+C)(2n-\rho+1)}
+ \frac{\delta(1-\rho)(1+\rho-2\delta)(B+D) + 4(B-D)n + 16(B-D)n^2}
{\rho(1-\rho)^2(A+C) + 4(A-C)n + 16(A-C)n^2}\right)\, .
\label{obs4}
\end{align}
To arrive at the above result, we use the horizon crossing
relation of $k$-th mode $k = aH$ to determine
$\frac{d|\tau|}{d\ln{k}} = -|\tau|$ i.e., the horizon exit time
$|\tau|$ increases as the momentum of the perturbation mode
decreases, as expected. Eq.~(\ref{obs4}) indicates that as similar
to $n_s$ and $r$, the running index ($\alpha$) also depends on the
parameters $R_h/a_0$ and $n$. Taking $R_h/a_0 = 0.05$, we give a
plot of $\alpha$ with respect to $n$ in Fig.~\ref{plot2}.
As it can be seen in Fig.~\ref{plot2}, the parameter $\alpha$
takes negative values, crossing  zero near about $n \simeq 0.30$.
Thus $\alpha$ lies within the Planck constraint for $0.30 \lesssim
n \lesssim 0.40$, which includes the matter bounce scenario. For
the Lagrange multiplier generalized $F(R)$ gravity model, we
showed that the pure matter bounce scenario as well as the quasi
matter bounce scenario are consistent with the Planck
observations. Therefore the generalized $F(R)$ gravity with the
Lagrange multiplier has a richer phenomenology in comparison to
scalar-tensor or standard $F(R)$ gravity model, which fails to
describe in a viable way these two { bouncing cosmology} scenarios.


\subsection{Stability of the Scalar and Tensor Perturbations}

As it is can be seen by Eqs.~(\ref{sp2}) and (\ref{tp2}), the
scalar and tensor perturbations are stable if the conditions
$z(t)^2 > 0$ and $z_T(t)^2 > 0$ are satisfied, respectively. Recall
that $z(t)$ and $z_T(t)$ have the following expressions:
\begin{align}
z(t)^2 = \frac{a(t)^2}{\left(H(t) + \frac{1}{2f'(R)}\frac{df'(R)}{dt}\right)^2}
\left[\frac{2E\sqrt{G}}{a^3} +
\frac{3}{2f'(R)}\left(\frac{df'(R)}{dt}\right)^2\right]\, ,
\label{sc1}
\end{align}
and
\begin{align}
z_T(t)^2 = a^2~f'(R)\, ,
\label{sc2}
\end{align}
as shown in Eqs.~(\ref{sp3}) and (\ref{tp3}). However, as we
mentioned earlier, the stability condition has to be checked for
all cosmic times, including the bouncing point which occurs at
$t=0$, where the low-curvature approximation no longer holds true.
Thus, it will not be justified if we use the form of $F(R)$
obtained in Eq.~(\ref{low curvature sol2}) to check the stability
near the bouncing point. On the other hand, if we find the form of
$F(R)$ in the large-curvature regime ($R/a_0 \gtrsim 1$), then
such form of $F(R)$ cannot be used to examine the stability away
from the bouncing point. Thereby, the best way to investigate the
stability condition is to determine the forms of $F(R)$ and $G(R)$
for the whole range of time  $-\infty < t < \infty$, and then use
such forms of $F(R)$, $G(R)$ in the expression of $z(t)$ and
$z_T(t)$. For this purpose, we solve Eqs.~(\ref{FRW eqn 2}) and
(\ref{FRW eqn3}) numerically, and proceed as follows; first we
analytically solve Eqs.~(\ref{FRW eqn 2}) and (\ref{FRW eqn3}) in
the large-curvature limit ($\frac{R}{a_0} \gtrsim 1$) to estimate
the boundary conditions necessary for the numerical solution.
Using such boundary conditions, we then solve the equations numerically.

\subsubsection{The Large Curvature Limit}

In the large-curvature limit (or equivalently the small cosmic
time limit), the scale factor in Eq.~(\ref{scale factor}) becomes,
\begin{align}
a(t) = 1 + a_0nt^2\, .
\label{sc3}
\end{align}
The corresponding Hubble rate and the Ricci scalar reads,
\begin{align}
H(t) = 2na_0t \, , \quad
R(t) = 12na_0\left[1 + (4n-3)a_0t^2\right]\, .
\label{sc4}
\end{align}
The above expression for the Ricci scalar can be easily inverted
to get the function $t = t(R)$, by which we determine the Hubble
rate, its first derivative and the differential operators
expressed in terms of $R$ (with the condition $\frac{R}{a_0}
\gtrsim 1$) as follows,
\begin{align}
H(R) = \sqrt{\frac{n(R - 12na_0)}{3(4n-3)}}\, , \quad
\dot{H}(R) = 2na_0 - \frac{(R - 12na_0)}{2(4n-3)}\, ,
\label{sc5}
\end{align}
\begin{align}
H\frac{d}{dt}=& 4na_0 (R - 12na_0)~\frac{d}{dR}\, ,\nonumber\\
 \frac{d^2}{dt^2}=& 24na_0^2~(4n - 3)~\frac{d}{dR}
+ 48na_0^2~(4n - 3)(R - 12na_0)~\frac{d^2}{dR^2}\, .
\label{sc6}
\end{align}
With the above expressions, the gravitational equations
Eqs.~(\ref{FRW eqn 2}) and (\ref{FRW eqn3}) become,
\begin{align}
12na_0 (12na_0 - R)J''(R) + 3\left[2na_0
 - \frac{(12na_0 - R)}{(3 - 4n)}\left(\frac{1}{2}
 - \frac{n}{3}\right)\right]J'(R) - \frac{J(R)}{2} = 0\, ,
\label{sc7}
\end{align}
and
\begin{align}
F(R)/2=&\left[2na_0 + \left(n - \frac{1}{2}\right)
\frac{(12na_0 - R)}{(3 - 4n)}\right]J'(R)
+ \left[8na_0 (12na_0 - R) + 24na_0^2(3 - 4n)\right]J''(R)\nonumber\\
&-48na_0^2 (3 - 4n) (12na_0 - R) J'''(R)\, ,
\label{sc8}
\end{align}
respectively, with $J(R) = F(R) + \frac{2E\sqrt{G(R)}}{a^3(R)}$.
The solution of Eq.~(\ref{sc7}) is given in terms of the confluent
hypergeometric function as follows,
\begin{align}
J(R) = d~\left(\frac{2R}{a_0}\right)^{3/2}~
U\left[-\frac{(3+2n)}{2(3-2n)}, \frac{5}{2}, -\frac{(3-2n)}{2(3-4n)}
+ \frac{(3-2n)R/a_0}{24n(3-4n)}\right]\, ,
\label{sc9}
\end{align}
where $d$ is an integration constant and has mass dimension [+2].
The asymptotic behavior of the confluent Hypergeometric function
is given by $U[a,b,x] \sim x^{-a}$ when $x$ is large and thus, in the large curvature
limit, the solution $J(R)$ becomes,
\begin{align}
J(R)
\sim d~2^{3/2}~\left[\frac{(3-2n)}{24n(3-4n)}\right]^{\frac{(3+2n)}{2(3-2n)}}~\left(
\frac{R}{a_0}\right)^{(6-2n)/(3-2n)}\, .
\label{sc10}
\end{align}
Since the mass dimension of the integration constant $d$ is [+2], without loss of generality
we can take $d$ as
$d = \frac{a_0}{2^{3/2}} \left[\frac{24n(3-4n)}{(3-2n)}\right]^{\frac{(3+2n)}{2(3-2n)}}$
(as $a_0$ also has a mass dimension [+2]),
which immediately leads to the form of $J(R)$ as,
\begin{align}
J(R) \sim a_0 \left(\frac{R}{a_0}\right)^{(6-2n)/(3-2n)}\, .
\label{sc11}
\end{align}
Consequently the form of $F(R)$ can be obtained from
Eq.~(\ref{sc8}), and is given by the following expression,
\begin{align}
F(R) \sim a_0
\frac{(1-2n)(6-2n)}{(3-2n)(3-4n)}~\left(\frac{R}{a_0}\right)^{(6-2n)/(3-2n)}\, .
\label{sc12}
\end{align}
Thus the effective form of $f(R)$ is expressed as follows,
\begin{align}
f(R)=F(R) + \frac{E\sqrt{G(R)}}{a^3(R)} = \frac{1}{2}\left[J(R) + F(R)\right]
\sim a_0~\left[1 + \frac{(1-2n)(6-2n)}{(3-2n)(3-4n)}\right]~\left(
\frac{R}{a_0}\right)^{(6-2n)/(3-2n)}\, .
\label{sc13}
\end{align}
Eqs.~(\ref{low curvature sol3}) and (\ref{sc13}) indicate that in
the low-curvature regime, $f(R)$ goes as $f(R) \propto R^{\rho}$,
and in the large-curvature regime $f(R) \propto
R^{(6-2n/(3-2n))}$. Recall, $\rho =
\frac{1}{4}\left[3-2n-\sqrt{1+4n(5+n)}\right]$ which is negative
for $n > 0.25$, and as shown earlier, the present model is
consistent with the Planck results for $0.27 \lesssim n \lesssim 0.4$,
hence $\rho$ is negative, in order to ensure the viability of the
model. Furthermore, $\frac{(6-2n)}{3-2n}$ which is the exponent in
the large-curvature expression of $f(R)$ is greater than unity.
Therefore, it is clear that in the low-curvature regime $f(R)$ is
proportional to an inverse power of Ricci scalar $\propto
R^{-|\rho|}$, while in the large-curvature limit, $f(R)$ is given
by a higher power of $R$. Such functional forms of $f(R)$ gravity
are used quite frequently in the literature, since they allow
unification of  early with late-time acceleration. However in the
present paper, we get such form of $f(R)$ in the context of a
symmetric bouncing Universe, where the scale factor evolves as
$a(t) = (a_0t^2 + 1)^n$.

\subsubsection{Numerical Study of the Stability of Perturbations}\label{numerical study}

\begin{figure}[!h]
\begin{center}
 \centering
 \includegraphics[width=3.5in,height=2.0in]{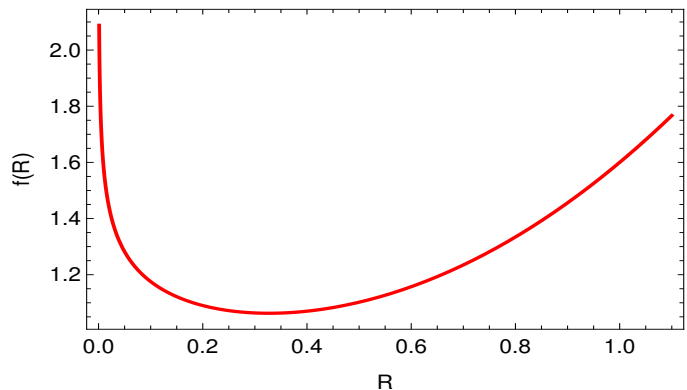}
 \caption{Numerical solution of $f = f(R)$ with $R$ being the independent variable. We take $a_0 = 1$ (in reduced Planck unit) and $n = 1/3$.}
 \label{plot_numerical}
\end{center}
\end{figure}

Now we proceed to the numerical solution study of the stability of
the perturbations. Using the above forms of $J(R)$ and $F(R)$ as
boundary conditions along with the expression $R(t) = 12n
\left[\frac{(4n-1)t^2 + 1/a_0}{\left(t^2 + 1/a_0\right)^2}\right]$, we
solve Eqs.~(\ref{FRW eqn 2}) and (\ref{FRW eqn3}) numerically,
with the cosmic time $t$ being the independent variable. Moreover,
$a_0$ and $n$ are taken as $a_0 = 1$ (in reduced Planck units) and
$n = 1/3$, respectively, so in effect we consider the matter bounce
scenario. However, it may be mentioned that the $n = 1/3$ case,
makes the model consistent with the Planck 2018 constraints, as
confirmed in the previous section. The numerical solution of
$f(R)$ in terms of $R$ is obtained by using the expression $R(t) =
12n \left[\frac{(4n-1)t^2 + 1/a_0}{\left(t^2 + 1/a_0\right)^2}\right]$
and is presented in Fig.~\ref{plot_numerical}.

It is evident from Fig.~\ref{plot_numerical} that $f(R)$ decreases
in the regime $R/a_0 \ll 1$, while in $R/a_0 \gtrsim 1$, $f(R)$
increases as a function of the Ricci scalar. This is expected from
the analytic solutions of $f(R)$ in the two limiting cases, see
Eqs.~(\ref{low curvature sol3}) and (\ref{sc13}). In the small and
large-curvature regimes, $f'(R)$ is given by
$\frac{-|\rho|}{R^{1+|\rho|}} < 0$ and
$\frac{(6-2n)}{(3-2n)}R^{3/(3-2n)} > 0$, respectively, which
justify the numerical solution of $f(R)$ in
Fig.~\ref{plot_numerical}.

\begin{figure}[!h]
\begin{center}
\centering
\includegraphics[width=3.0in,height=2.0in]{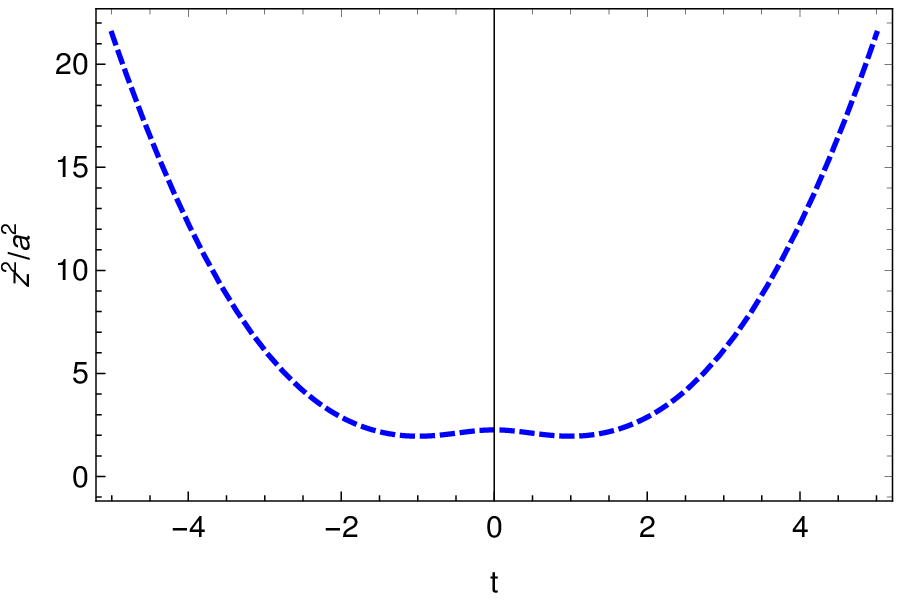}
\includegraphics[width=3.0in,height=2.0in]{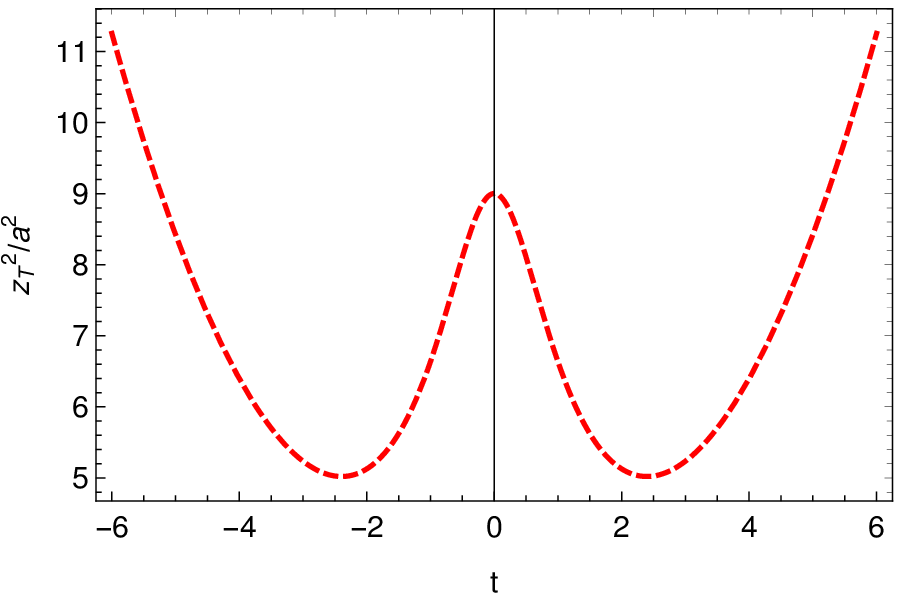}
\caption{$Left~part$ : $z^2/a^2$ vs. $t$ for the purpose of the stability of
the scalar perturbation. $Right~part$ : $z_T^2/a^2$ vs.$t$ for the
 purpose of stability of the tensor perturbation. In both cases, we take $a_0 = 1$
(in the reduced Planck unit) and $n = 1/3$.}
 \label{plot_perturbation}
\end{center}
\end{figure}

By plugging this numerical solution of $f(R)$ into the expressions
of $z(t)^2$ and $z_T(t)^2$, we can check the stability condition
of the metric perturbations for a wide range of the cosmic time.
It may be noticed that $z(t)^2$ and $z_T(t)^2$ carry a common
factor $a(t)^2$ which is always positive. Thus the stability
condition of the scalar and tensor perturbations are given by
$z(t)^2/a(t)^2 > 0$ and $z_T(t)^2/a(t)^2 > 0$, respectively. Using
the numerical solution shown in Fig.~\ref{plot_numerical}, we give
plots for $z(t)^2/a(t)^2$ and $z_T(t)^2/a(t)^2$ (with respect to
the cosmic time) in the left and right plots of
Fig.~\ref{plot_perturbation}, respectively, where we take $a_0 = 1$ and
$n = 1/3$. From Fig.~\ref{plot_perturbation} it is evident that
both the scalar and tensor perturbations are stable in the present
context. Moreover, as we mentioned earlier, the squared speed of
the perturbations are unity (i.e., $c_s^2 = c_T^2 = 1$) which
guarantees the absence of any ghost modes from the present model.
Thus for the matter bounce scenario materialized with the Lagrange
multiplier $F(R)$ gravity model, there exist a  range of the free
parameters, for which the model becomes compatible with the latest
Planck 2018 observations, and also becomes free from instabilities
of the metric perturbations.


\section{Energy condition}\label{sec_energy}

\begin{figure}[!h]
\begin{center}
 \centering
 \includegraphics[width=3.0in,height=2.0in]{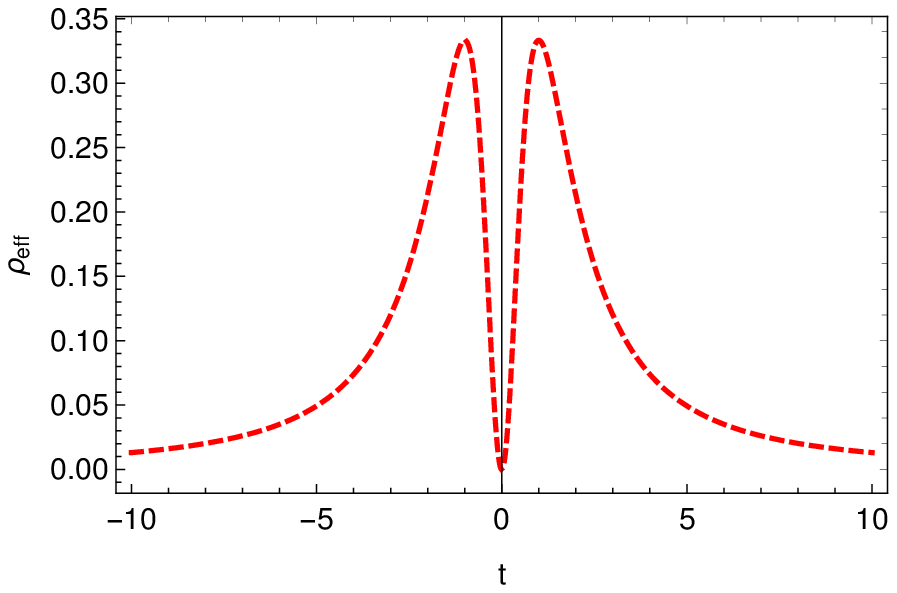}
 \includegraphics[width=3.0in,height=2.0in]{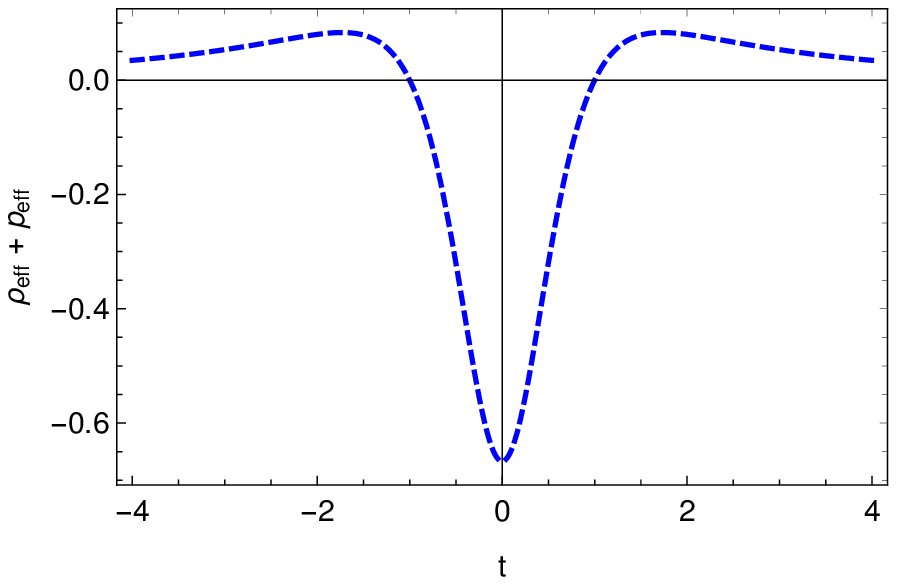}
 \caption{$Left~part$ : $\rho_\mathrm{eff}$ vs. $t$ for the purpose of weak energy condition. $Right~part$ : $\rho_\mathrm{eff} + p_\mathrm{eff}$ vs.$t$ for the
 purpose of null energy condition. In both cases, we take $E = 1$, $a_0 = 1$ (in reduced Planck
units) and $n = 1/3$.}
 \label{plot_energy}
\end{center}
\end{figure}

A crucial drawback in most of the bouncing models is the violation
of the null energy condition, which is also a vital ingredient of
the Hawking-Penrose theorems, in the context of Einstein's general
relativity. Here we check the energy conditions in the context of the
Lagrange multiplier $F(R)$ gravity model. For this purpose, we
determine the effective energy density $\rho_\mathrm{eff}$ and pressure
$p_\mathrm{eff}$ from Eqs.~(\ref{FRW eqn 2}) and (\ref{FRW eqn3}), as
follows,
\begin{align}
\rho_\mathrm{eff}=&\frac{1}{2\left(F'(R) + \frac{EG'}{a^3\sqrt{G}}\right)}
\left[F(R)-\frac{E\sqrt{G}}{a^3}+3\left(\frac{d^2}{dt^2}+H\frac{d}{dt}\right)
\left(F'(R) + \frac{EG'}{a^3\sqrt{G}}\right)\right]\nonumber\, ,\\
\rho_\mathrm{eff} + p_\mathrm{eff}=&-\frac{1}{\left(F'(R)
+ \frac{EG'}{a^3\sqrt{G}}\right)}\left[\frac{E\sqrt{G}}{a^3}
+\left(-\frac{d^2}{dt^2}+H\frac{d}{dt}\right)
\left(F'(R) + \frac{EG'}{a^3\sqrt{G}}\right)\right]\, .
\label{enenrgy1}
\end{align}
By using the above expressions along with the numerical solution
of $f(R)$ determined in the previous section, we give the plots of
$\rho_\mathrm{eff}$ and $\rho_\mathrm{eff} + p_\mathrm{eff}$ (with respect to cosmic
time) in the left and right plots of Fig.~\ref{plot_energy},
respectively, for $n = 1/3$, $E = 1$, $a_0 = 1$ (in reduced Planck
units).
As it can be seen in Fig.~\ref{plot_energy}, $\rho_\mathrm{eff}$
remains positive for the whole time regime (or equal to zero at
the bouncing point), while $\rho_\mathrm{eff} + p_\mathrm{eff}$
becomes negative near the bouncing point. This indicates that the
null energy condition is violated, which further implies that the
weak energy condition is necessarily violated. At this stage, we
want to mention that the holonomy-corrected generalized $F(R)$
gravity model or the presence of extra spatial dimension, where
$H^2$ is proportional to linear powers, as well as quadratic
powers of the energy density, may play a significant role to
rescue the null energy condition for a non-singular
bounce.\\
However before moving to the next section, we want to state that
the present paper studies `quasi-matter bounce' in Lagrange
multiplier F(R) gravity model, which is found to yield a nearly
scale-invariant power spectra of scalar and tensor perturbations
adiabatically. However, at the background level of the contracting
era, it is also known that matter (or quasi-matter) contraction is
not an attractor, and worse, it is unstable to the growth of
anisotropies (more explicitly the anisotropy grows with the scale
factor as $1/a^6$ which is known as the BKL instability; see
\cite{new1,Levy:2016xcl}). Thus the present model remains at the
level of a toy model with this respect. However in the ekpyrotic
bounce scenario (instead of matter or quasi-matter bounce)
\cite{Cai:2013vm,Erickson:2003zm,Garfinkle:2008ei}, the BKL
instability does not occur and it will be an interesting avenue to
explore the possible effects of Lagrange multiplier term in an
ekpyrotic bounce scenario, which is expected to study in a near
future work.\\

\section{Standard $F(R)$ gravity and the comparison with Lagrange Multiplier $F(R)$ gravity}\label{sec_standard_F(R)}

\begin{figure}[!h]
\begin{center}
 \centering
 \includegraphics[width=3.5in,height=2.0in]{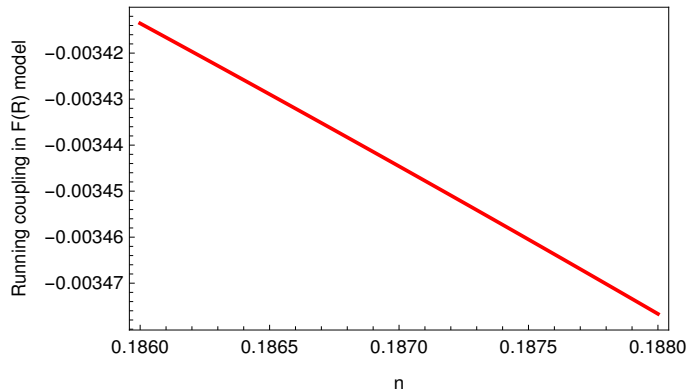}
 \caption{Parametric plot of $\tilde{\alpha}$ vs. $n$ for
$R_h/a_0 = 2\times 10^{-4}$.}
 \label{malakies1}
\end{center}
\end{figure}

\begin{figure}[!h]
\begin{center}
 \centering
 \includegraphics[width=3.5in,height=2.0in]{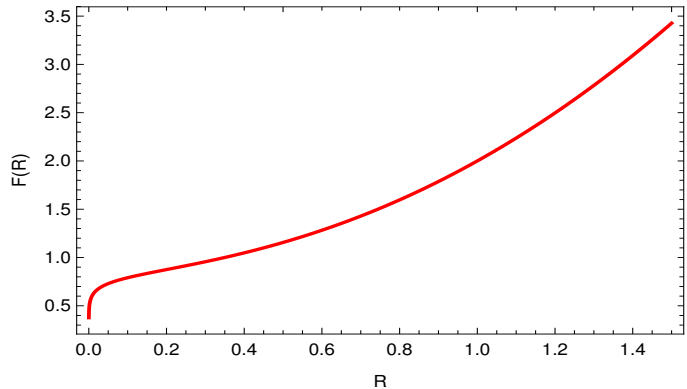}
 \caption{Numerical solution of F(R) with $n = 0.186$ and $a_0 = 1$}
 \label{plot_numerical_F(R)}
\end{center}
\end{figure}

\begin{figure}[!h]
\begin{center}
 \centering
 \includegraphics[width=3.0in,height=2.0in]{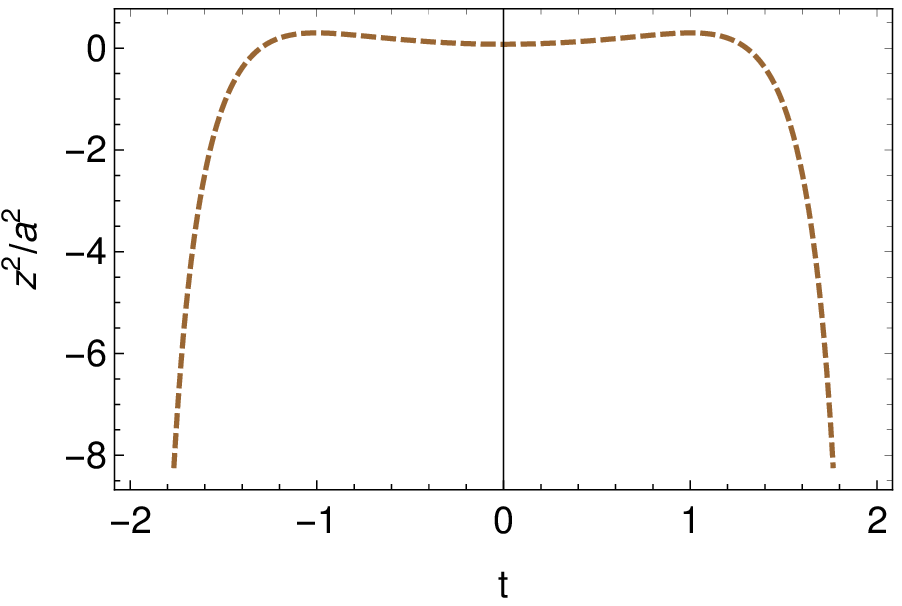}
 \includegraphics[width=3.0in,height=2.0in]{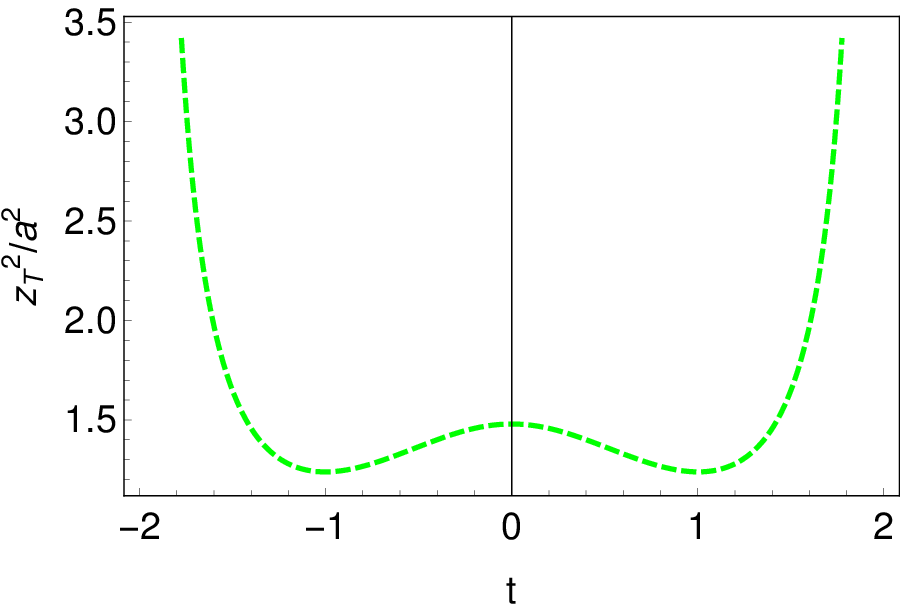}
 \caption{$Left~part$ : $z^2/a^2$ vs. $t$ for the purpose of the stability of the scalar perturbation. $Right~part$ : $z_T^2/a^2$ vs.$t$ for the
 purpose of stability of the tensor perturbation. In both cases, we take $a_0 = 1$ (in reduced Planck units) and $n = 0.186$.}
 \label{plot_perturbation_F(R)}
\end{center}
\end{figure}

In this section, we consider the standard $F(R)$ gravity to study
the realization of the bouncing universe of Eq.~(\ref{scale
factor}) and we shall compare the results with those obtained for
the Lagrange multiplier $F(R)$ gravity model. The action for
vacuum $F(R)$ gravity model is given by,
\begin{align}
S = \frac{1}{2\kappa^2} \int d^4x \sqrt{-g} F(R) \, ,
\label{f1}
\end{align}
where $\frac{1}{\kappa^2}=M^2$ with $M$ being again the four
dimensional Planck mass. The action in Eq.~(\ref{f1}) leads to the
following Friedmann equations of motion,
\begin{align}
 -\frac{F(R)}{2} + 3\left(\dot{H} + H^2\right)F'(R) - 3H\frac{dF'}{dt} = 0\, ,\nonumber\\
\frac{F(R)}{2} - \left(\dot{H} + 3H^2\right)F'(R) + \left(\frac{d^2}{dt^2}
+ 2H\frac{d}{dt}\right)F'(R) = 0\, ,
\label{f2}
\end{align}
with $H$ being the Hubble rate. We should note that
Eqs.~(\ref{FRW eqn 2}) and (\ref{FRW eqn3}) become identical with
Eq.~(\ref{f2}) by choosing $E = 0$, since for $E = 0$, the Lagrange
multiplier field $\lambda(t)$ goes to zero and thus the
gravitational equations of a Lagrange multiplier $F(R)$ gravity
model become identical with that of a standard $F(R)$ gravity. As we
discussed earlier, for the purpose of determining the observable
quantities, the low-curvature limit i.e., $R/a_0 \ll 1$ is a viable
approximation. Recall that, in the regime $R/a_0 \ll 1$, the Ricci
scalar can be written as $R(t) = \frac{12n(4n-1)}{t^2}$. This
along with the expressions of the Hubble rate, its first
derivative and the differential operators, as determined in
Eq.~(\ref{low curvature quantites}), will enable us to express the
gravitational equations as follows,
\begin{align}
\frac{2}{(4n-1)}~R^2F''(R) - \frac{(1-2n)}{(4n-1)}~RF'(R) - F(R) = 0 \, ,
\label{f3}
\end{align}
which can be solved as,
\begin{align}
F(R) = a_0 \left[\left(\frac{R}{a_0}\right)^{\rho}
+ \left(\frac{R}{a_0}\right)^{\delta}\right] \, ,
\label{f4}
\end{align}
with $\rho = \frac{1}{4}\left[3-2n-\sqrt{1+4n(5+n)}\right]$ and
$\delta = \frac{1}{4}\left[3-2n+\sqrt{1+4n(5+n)}\right]$.
Eq.~(\ref{f4}) represents the reconstructed form of $F(R)$ gravity
for the bouncing scale factor $a(t) = (a_0t^2 + 1)^n$. This form
of F(R) matches with the reconstructed form of effective f(R) (see
Eq.(\ref{low curvature sol3}), apart from the coefficients) in
Lagrange multiplier
higher curvature model in the low curvature regime.\\
Before moving towards the perturbation, at this stage we want to
study whether this form of $F(R)$ (in Eq. (\ref{f4})) passes the
astrophysical tests in low curvature regime.  An example of the
tests is matter instability, which is related to the fact that the
spherical body solution in general relativity may not be the
solution in modified gravity theory. The matter instability may
appear when the energy density or the curvature is large compared
with the average density or curvature in the universe, as is the
case inside of a planet. Following \cite{Nojiri:2010wj}, we
immediately write the potential ($U(R_b)$, with $R_b$ be the
perturbed Ricci scalar) for the perturbed Ricci curvature over
Einstein gravity as,
\begin{eqnarray}
 U(R_b) = \frac{R_b}{3} - \frac{F^{(1)}(R_b)F^{(3)}(R_b)R_b}{3F^{(2)}(R_b)^2} - \frac{F^{(1)}(R_b)}{3F^{(2)}(R_b)} +
 \frac{2F(R_b)F^{(3)}(R_b)}{3F^{(2)}(R_b)^2} - \frac{F^{(3)}(R_b)R_b}{3F^{(2)}(R_b)^2}
 \label{astro_test1}
\end{eqnarray}
where we denote $d^kF(R)/dR^k = F^{(k)}(R)$. If $U(R_b) > 0$, the
perturbation grows with time and the system becomes unstable.
Recall $\rho < 0$ and $\delta > 0$ and thus the term $R^{\rho}$
dominates over $R^{\delta}$ in the low curvature regime. Thus we
can approximate $F(R) \sim R^{\rho}$ in the low curvature regime
which immediately leads to the potential as,
\begin{eqnarray}
 U(R_b) = -\frac{2(|\rho| + 2)}{9|\rho|(|\rho| + 1)}R_b + \frac{(|\rho| + 2)}{3|\rho|(|\rho| + 1)}R_b^{2+|\rho|}
 \label{astro_test2}
\end{eqnarray}
In the low curvature regime, the first term dominates in the above expression of $U(R_b)$ and thus $U(R_b)$ becomes less than zero. This indicates that
the model considered here passes the matter instability test. However more checks of this theory should be done in order to conclude if
the model is realistic one or not, which we expect to study in a future work.\\
Having the reconstructed form of
$F(R)$ in hand, we proceed to study the cosmological perturbations
in this model and the perturbed metric is given by,
\begin{align}
ds^2 = -\left(1 + 2\tilde{\Psi}\right)dt^2 + a(t)^2\left(1 - 2\tilde{\Psi}\right)
\left(\delta_{ij} + \tilde{h}_{ij}\right)~dx^idx^j\, ,
\label{f5}
\end{align}
where $\tilde{\Psi}(t,\vec{x})$ and $\tilde{h}_{ij}(t,\vec{x})$
are scalar and tensor perturbed variable, respectively. The
``tilde'' quantities are reserved for the pure $F(R)$ gravity
model, in order to make a comparison with the Lagrange multiplier
$F(R)$ gravity model. Using the same procedure as discussed in
section \ref{sec_perturbation}, we obtain the first order
perturbed equations in $F(R)$ gravity model as follows,
\begin{align}
\frac{d^2\tilde{v}}{d\tau^2} + \left[k^2 - \frac{1}{\tilde{z}}
\frac{d^2\tilde{z}}{d\tau^2}\right]\tilde{v}(\tau)=0\, ,\quad
\frac{d^2\tilde{v}_T}{d\tau^2} + \left[k^2 - \frac{1}{\tilde{z}_T}
\frac{d^2\tilde{z}_T}{d\tau^2}\right]\tilde{v}_T(\tau)=0\, ,
\label{f6}
\end{align}
where $\tau$ is the conformal time given by $\tau =
\frac{t^{1-2n}}{a_0^n~(1-2n)}$. Moreover $\tilde{z}(\tau)$,
$\tilde{z}_T(\tau)$ are the scalar, tensor type Mukhanov-Sasaki
variable, respectively, and have the following forms,
\begin{align}
\tilde{z}[\tau(t)]=\frac{a(t)}{\left(H(t)
+ \frac{1}{2F'(R)}\frac{dF'(R)}{dt}\right)} \sqrt{\frac{3}{2F'(R)}
\left(\frac{dF'(R)}{dt}\right)^2}\, ,\quad
\tilde{z}_T[\tau(t)]=a\sqrt{F'(R)}\, .
\label{f7}
\end{align}
Comparing the above expression with Eqs.~(\ref{sp3}) and
(\ref{tp3}), it is clearly observed that for $E = 0$, the scalar
and tensor type Mukhanov-Sasaki variables in a standard $F(R)$
gravity model become same with that of the Lagrange multiplier
$F(R)$ gravity model, as expected. Using Eq.~(\ref{f7}), we
further obtain the following expressions (in the low-curvature
regime) which are important towards solving the Mukhanov
equations,
\begin{align}
\frac{1}{\tilde{z}}\frac{d^2\tilde{z}}{d\tau^2}
= \frac{\xi(\xi-1)}{\tau^2} \left[1 + \frac{2(\delta-\rho)}{(\xi-1)}
\left(R/a_0\right)^{\delta-\rho}~\left(\frac{\delta(\rho-\delta)}{\rho(2n-\rho+1)}
+ \frac{\delta(1+\rho-2\delta)}{\rho(1-\rho)}\right)\right]\, ,
\label{f8}
\end{align}
and
\begin{align}
\frac{1}{\tilde{z}_T}\frac{d^2\tilde{z}_T}{d\tau^2} = \frac{\xi(\xi-1)}{\tau^2}
\left[1 - \frac{2\delta(\delta-\rho)}{(\xi-1)\rho}
\left(R/a_0\right)^{\delta-\rho}\right]\, ,
\label{f9}
\end{align}
and recall that $\xi = \frac{(2n+1-\rho)}{(1-2n)}$. As
$\delta-\rho$ is a positive quantity, the terms in the parenthesis
in Eqs.~(\ref{f8}) and (\ref{f9}), can be safely considered to be
small in the low-curvature regime and consequently
$\tilde{z}''(\tau)/\tilde{z}$ and
$\tilde{z}_T''(\tau)/\tilde{z}_T$ become proportional to
$1/\tau^2$. In effect, the solutions of the Mukhanov variables are
expressed in terms of the Hankel function as discussed earlier in
section \ref{sec_perturbation}. With these solutions, we obtain
the spectral index $\tilde{n}_s$ and the tensor-to-scalar ratio
$\tilde{r}$ for the standard $F(R)$ gravity model as follows,
\begin{align}
\tilde{n}_s = 4 - \sqrt{1 + 4\xi(\xi-1)\left[1
+ \frac{2(\delta-\rho)}{(\xi-1)}~\left(R_h/a_0
\right)^{\delta-\rho}~\left(\frac{\delta(\rho-\delta)}{\rho(2n-\rho+1)}
+ \frac{\delta(1+\rho-2\delta)}{\rho(1-\rho)}\right)\right]}\, ,
\label{f10}
\end{align}
and
\begin{align}
\tilde{r} = \left. 2\left[\frac{\tilde{z}(\tau)}{\tilde{z}_T(\tau)}\right]^2
\right|_{\tau = \tau_h}
 = \left. 3\left[\frac{1}{F'(R)\left(H(t) + \frac{1}{2F'(R)}\frac{dF'(R)}{dt}\right)}
\right]^2~\left(\frac{dF'(R)}{dt}\right)^2
\right|_{t = t_h}\, ,
\label{f11}
\end{align}
respectively, where $t_h$ is the time of horizon exit and $R_h =
R(t_h)$. Similarly to the previously discussed Lagrange multiplier
$F(R)$ gravity model, see section \ref{sec_perturbation}, the
spectral index and tensor to scalar ratio of the $F(R)$ gravity
model depend on the parameters $R_h/a_0$ and $n$ (both are
dimensionless parameters). With these expressions, we can confront
the observational parameters of the models with the Planck 2018
results. For the $F(R)$ gravity model, the spectral index
($\tilde{n}_s$) lies within the Planck constraints for a narrow regime
of the parameters as: $10^{-4} \lesssim \frac{R_h}{a_0} \lesssim
3\times 10^{-4}$ and $0.1860 \leq n \leq 0.1866$. However for these
values of the free parameters, the tensor-to-scalar ratio takes
values in the range $1.9915 \lesssim \tilde{r} \lesssim 1.9940$
and hence is not compatible with the Planck results. Thereby, we
can argue that $\tilde{n}_s$ ad $\tilde{r}$ are not simultaneously
compatible with the Planck constraints for a bouncing universe
(with $a(t) = (a_0t^2 + 1)^n$) in the standard $F(R)$ model, in
contrast to the Lagrange multiplier $F(R)$ gravity model. This
clearly indicates the importance of the Lagrange multiplier field
$\lambda(t)$, present in action (\ref{action1}) in making the
compatibility of the observational parameters with the Planck
results. Furthermore the running of the spectral index is
determined as,
\begin{align}
\tilde{\alpha} = \left. \frac{dn_s}{d\ln{k}} \right|_{\tau=\tau_h}
= \frac{4\xi(\delta-\rho)^2}{\sqrt{1 + 4\xi(\xi-1)}}\left(\frac{R_h}{a_0}
\right)^{\delta-\rho}~\left(\frac{\delta(\rho-\delta)}{\rho(2n-\rho+1)}
+ \frac{\delta(1+\rho-2\delta)}{\rho(1-\rho)}\right) \, ,
\label{f12}
\end{align}
where we used the relation of horizon crossing of the $k$-th mode,
that is $k = aH$. It turns out that $\tilde{\alpha}$ lies within
the Planck constraints $\alpha = -0.0085 \pm 0.0073$ for $0.1860
\leq n \leq 0.1866$ along with $R_h/a_0 = 2\times 10^{-4}$ (the
regime where $\tilde{n}_s$ is also compatible with the Planck
results) and this is demonstrated in Fig.~\ref{malakies1}.
Thus in conclusion, in the $F(R)$ model, the spectral index and
the running index are consistent with the Planck results while the
tensor to scalar ratio is not. However, the presence of the field
$\lambda(t)$ in the $F(R)$ gravity model, makes all the three
parameters simultaneously compatible with observational
constraints. This makes clear that the field $\lambda(t)$ has a
significant contribution on the observational parameters. Next we
proceed to explore the stability condition of the $F(R)$ model. In
order to investigate the stability condition, we need to determine
the form of $F(R)$ for the whole duration of the bounce, and for
this purpose we solve Eq.~(\ref{f2}) numerically. However, before
going to the numerical solution, we reconstruct the form of $F(R)$
for $a(t) = (a_0t^2 + 1)^n$ in the large-curvature regime $R/a_0
\gtrsim 1$, which will act as boundary condition in determining
the numerical solution. In the large curvature regime the $F(R)$
gravitational equation becomes,
\begin{align}
12na_0 (12na_0 - R)F''(R) + 3\left[2na_0
 - \frac{(12na_0 - R)}{(3 - 4n)}\left(\frac{1}{2}
 - \frac{n}{3}\right)\right]F'(R) - \frac{F(R)}{2} = 0\, .
\label{f13}
\end{align}
The solution of Eq.~(\ref{f13}) is given in terms of the
Hypergeometric function, and by using the asymptotic behavior of
the Hypergeometric function, we can write the solution of $F(R)$
in regime $R/a_0 \gtrsim 1$ as follows,
\begin{align}
F(R) \sim a_0 \left(\frac{R}{a_0}\right)^{(6-2n)/(3-2n)} \, .
\label{f14}
\end{align}
By using the above expression as a boundary condition along with
$a_0 = 1$, $n = 0.186$, we obtain the numerical solution of $F(R)$
from Eq.~(\ref{f2}). This is depicted in
Fig.~\ref{plot_numerical_F(R)}.
As it can be seen in Fig.~\ref{plot_numerical_F(R)}, the $F(R)$
starts from zero (at $R \sim 0$) and gradually increases with the
Ricci scalar, unlike to the case of the Lagrange multiplier $F(R)$
gravity model, where $f(R)$ actually diverges at $R = 0$ (see
Fig.~\ref{plot_numerical}). This feature occurs due to the different
viability regime of the parameter $n$, which makes the
corresponding model, consistent with the Planck results. In the
Lagrange multiplier $F(R)$ gravity model, the viability range of
$n$ is given by $0.27 \lesssim n \lesssim 0.40$, which makes
$\rho$ $\left(= \frac{1}{4}\left[3-2n-\sqrt{1+4n(5+n)}\right]\right)$ a negative
quantity. Thus the effective $f(R)$ behaves as an inverse power of
$R$ in the low-curvature regime and diverges at $R = 0$, as shown
in Fig.~\ref{plot_numerical}. On the other hand, for the standard
$F(R)$ model the viability regime of $n$, in terms of spectral
index and running index, is given by $0.1860 \lesssim n \lesssim
0.1866$ which makes $\rho$ a positive quantity. As a result $F(R)$
behaves as a positive power of $R$ in the low-curvature regime and
goes to zero at $R = 0$, as depicted in
Fig.~\ref{plot_numerical_F(R)}. By using the numerical solution of
$F(R)$, we give the plots of $\tilde{z}^2/a^2$ and
$\tilde{z}_T^2/a^2$ (with respect to time) in the left and right
plots of Fig.~\ref{plot_perturbation_F(R)} to check the stability
of the scalar ($\tilde{\Psi}$) and tensor ($\tilde{h}_{ij}$)
perturbation, respectively.
Fig.~{\ref{plot_perturbation_F(R)}} clearly reveals that, for the
$F(R)$ model at hand, the scalar perturbation is not stable while
the tensor perturbation is, in contrast to the case of the
generalized Lagrange multiplier $F(R)$ gravity, where both the
scalar and tensor perturbations are found to be stable, see
Fig.~\ref{plot_perturbation}. Actually, the absence of the Lagrange
multiplier field destabilizes the scalar perturbations without
affecting though the stability of the tensor perturbations.
Regarding the energy conditions in the $F(R)$ model, it turns out
that the weak energy condition is satisfied while the null energy
condition is violated near the bouncing point, for the bouncing
universe described by $a(t) = (a_0t^2 + 1)^n$. The comparison of
the standard vacuum $F(R)$ gravity with that of the Lagrange
multiplier $F(R)$ gravity is shown in Table \ref{table}.
\begin{table}[h]
\centering
\begin{tabular}{|c| c| c|}
 \hline
Observable quantities, stability the energy conditions &
the Lagrange multiplier $F(R)$ model & Standard $F(R)$ model \\
\hline\hline
1. Observable quantities & Viable & Not Viable\\
\hline
2. Scalar perturbation & Stable  & Not stable\\
\hline
3. Tensor perturbation & Stable  & Stable \\
\hline
4. Weak energy condition & Violated & Violated\\
\hline
5. Null energy condition & Violated & Violated\\
\hline
\end{tabular}
\caption{Comparison of observable quantities, stability of the perturbations
and the energy conditions
for the Lagrange multiplier $F(R)$ model and the standard $F(R)$ gravity}
\label{table}
\end{table}

Thus the bouncing universe with $a(t) = (a_0t^2 + 1)^n$ is well
described by the Lagrange multiplier $F(R)$ model in comparison to
the standard $F(R)$ gravity. However the presence of the Lagrange
multiplier field cannot rescue the null energy condition. In this
regard, we want to mention that the holonomy corrected generalized
$F(R)$ gravity where $H^2$ is proportional to linear as well as
squared of the effective energy density, may rescue the energy
condition and we defer this task to a near future work.


\section{Comoving Hubble radius : Viability of the Low-curvature Limit}

Before concluding, let us comment on an interesting issue, related
to the viability of the low-curvature approximation that we have
considered in calculating the scalar and tensor power spectrum in
section \ref{sec_perturbation}. In the context of the matter
{ bounce cosmology}, which is obtained by taking $n=1/3$ in
Eq.~(\ref{scale factor}), the primordial perturbations of the comoving
curvature, which originate from quantum vacuum fluctuations, were
at subhorizon scales during the contracting era in the
low-curvature regime, that is, their wavelength was much smaller
than the comoving Hubble radius which is defined by $r_h =
\frac{1}{aH}$. In the matter bounce evolution, the Hubble horizon
radius decreases in size, and this causes the perturbation modes
to exit from the horizon eventually, with this exit occurring when
the contracting Hubble horizon becomes equal to the wavelength of
these primordial modes. However, in the present context, we
consider a larger class of bouncing models of the form $a(t) =
(a_0t^2 + 1)^n$, in the presence of a generalized Lagrange
multiplier $F(R)$ gravity. In such higher curvature model, it
turns out that the observable quantities lie within the Planck
constraints when the parameter values are taken in the range $0.01
\lesssim \frac{R_h}{a_0} \lesssim 0.07$ and $0.27 \lesssim n
\lesssim 0.40$ and moreover, by calculating the observable
quantities, we have assumed that the horizon exit of the
perturbation modes occurred during the low-curvature regime of the
contracting era. Thus, it will be important to check what are the
possible values of $n$ which make the low-curvature limit a viable
approximation in calculating the power spectrum for the bouncing
model $a(t) = (a_0t^2 + 1)^n$.

\begin{figure}[!h]
\begin{center}
 \centering
 \includegraphics[width=3.5in,height=2.0in]{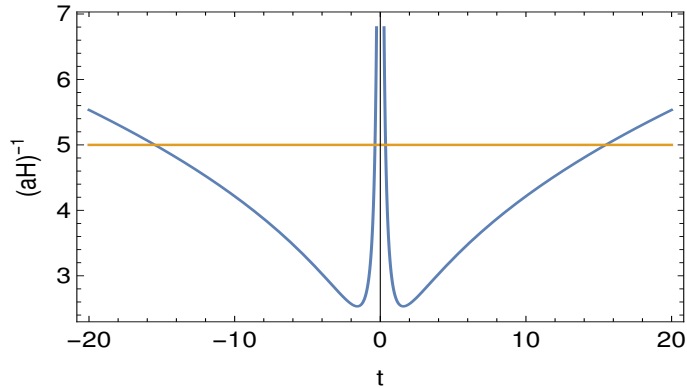}
 \caption{Comoving Hubble radius (Blue curve) and a perturbation mode (yellow curve) with respect to cosmic time for $n = 0.30$}
 \label{plot_horizon}
\end{center}
\end{figure}

The expression of the scale factor in Eq.~(\ref{scale factor})
immediately leads to the Comoving Hubble radius,
\begin{align}
r_h = \frac{(1 + a_0t^2)^{1-n}}{2a_0nt} \, .
\label{viability1}
\end{align}
Thereby $r_h$ diverges at $t \simeq 0$, as expected because the
Hubble rate goes to zero at the bouncing point. Furthermore,
the asymptotic behavior of $r_h$ is given by $r_h \sim t^{1-2n}$,
thus $r_h\left(|t|\rightarrow \infty\right)$ diverges for $n < 1/2$,
otherwise $r_h$ goes to zero asymptotically. Hence, for $n < 1/2$,
the comoving Hubble radius decreases initially in the contracting
era and then diverges near the bouncing point, unlike to the case
$n > 1/2$ where the Hubble radius increases from the infinite past
and gradually diverges at $t = 0$. As a result, the possible range
of $n$ which leads the perturbation modes to exit the horizon at
large negative time and make the low-curvature limit a viable
approximation in calculating the power spectrum, is given by $0 <
n < 1/2$. Moreover this range of $n$ also supports the range $0.27
\lesssim n \lesssim 0.40$, which makes the observable quantities
simultaneously compatible with the Planck 2018 results. In
Fig.~\ref{plot_horizon} we plot the comoving Hubble radius and a
perturbation mode as functions of the cosmic time for $n = 0.30$.

\section{Conclusions}\label{sec_conclusion}

In this paper, we considered a variant matter { bounce cosmology}
with $a(t) = (a_0t^2 + 1)^n$, in the context of the Lagrange
multiplier $F(R)$ gravity model. For such model, it was shown that
for $n < 1/2$, the perturbation modes, which were generated during
the contacting era, exit from the comoving Hubble radius at large
negative time, deeply in the contracting era, which in turn makes
the low-curvature limit a viable approximation in calculating the
observable quantities. Thus we constructed the form of effective
$f(R)$ gravity that may materialize the above cosmic scenario, and
consequently we determined the scalar and tensor power spectrums
in the low-curvature regime. These lead to the expressions of
various observable quantities like spectral index of primordial
scalar perturbations, the tensor-to-scalar ratio and the running
of spectral index, which were found to depend on the dimensionless
parameters $R_h/a_0$ and $n$, with $R_h$ being the Ricci curvature
at horizon exit. It turned out that such observable quantities are
simultaneously compatible with the Planck 2018 constraints for the
parameters chosen in the range
 $0.01 \lesssim \frac{R_h}{a_0} \lesssim 0.07$ and $0.27
\lesssim n \lesssim 0.40$. It may be noticed that this range of
$n$ is supported by the range $0 < n < 1/2$ which makes the
low-curvature approximation,  a valid approximation in calculating
the power spectrums. The stability condition of the metric
perturbations had to be checked for all cosmic times, include the
bouncing point time instance $t = 0$, where the low-curvature
approximation does not hold true. For the purpose of examining the
stability conditions, we determined the $f(R)$ gravity beyond the
low-curvature regime, in particular, for the whole range of time
($-\infty < t  < \infty$) by solving the Friedmann equations
without the low-curvature approximation, however numerically. The
numerical solution clearly depicted that in the low-curvature
regime, $f(R)$ is proportional to an inverse power of the Ricci
scalar, while in the large curvature limit, $f(R)$ is given by a
higher power (higher than one) of $R$. Such characteristics of
$f(R)$ gravity have been widely used in the literature mainly in
order to unify the early-time with the late-time acceleration.

By using the obtained numerical solutions, we checked the
stability conditions and as a result we found that both the scalar
and tensor perturbations were stable for the same range of
parameter values, which guaranteed the phenomenological viability
of the model. Moreover the Mukhanov-Sasaki equations suggest that
the squared speed of the gravity waves is unity, which confirmed
the absence of any ghost modes. We further calculated the
effective energy density and pressure in the present context in
order to investigate the energy conditions. As a consequence, we
found that both the null energy and the weak energy conditions are
violated near the bouncing point. The phenomenology of the present
non-singular bounce is also discussed in the context of a standard
$F(R)$ gravity model. We found that the results obtained in the
Lagrange multiplier $F(R)$ gravity model, are in contrast with the
standard $F(R)$ model for which the spectral index and the running
index are simultaneously compatible with the Planck results,
however the tensor-to-scalar ratio is not. Also the scalar
perturbation is stable but the model is plagued with the
instability of the tensor perturbation and finally, the weak
energy condition is satisfied while the null energy condition is
violated.

The above features clearly justify the importance of the Lagrange
multiplier field in making the observational indices compatible
with the Planck data and also in removing the instability of the
metric perturbations. Therefore the bouncing universe with $a(t) =
(a_0t^2 + 1)^n$ is well described by the Lagrange multiplier
$F(R)$ gravity model in comparison to the standard $F(R)$ model.
However the presence of the Lagrange multiplier field cannot evade
the violation of null energy condition. In this regard, we want to
mention that the holonomy-corrected higher curvature model may
rescue the null energy condition and we hope to address this issue
in a future work.

\section*{Acknowledgments}

This work is supported by MINECO (Spain), FIS2016-76363-P, and by
project 2017 SGR247 (AGAUR, Catalonia) (S.D.O). This work is also
supported by MEXT KAKENHI Grant-in-Aid for Scientific Research on
Innovative Areas ``Cosmic Acceleration'' No. 15H05890 (S.N.) and
the JSPS Grant-in-Aid for Scientific Research (C) No. 18K03615
(S.N.).

\end{document}